\documentclass[12pt]{article}

\usepackage[margin=1in]{geometry}
\usepackage{amsmath, amssymb, amsfonts}
\usepackage{graphicx}
\usepackage{booktabs}
\usepackage{natbib}
\usepackage{setspace}
\usepackage{hyperref}
\usepackage{xcolor}
\usepackage{graphicx}
\usepackage{amsthm}
\usepackage{multirow}
\usepackage{placeins}
\usepackage{bm}\usepackage{caption}
\newtheorem{proposition}{Proposition}

\title{An Analysis of High-Dimensional Land-Water Transitions in the Great Lakes via a Bayesian Discrepancy Model for Multinomial Data}

\author{
Seulki Lee\thanks{\texttt{sl8b6@missouri.edu}}
\qquad
Jonathan R. Bradley\thanks{Corresponding author: \texttt{bradleyjr@missouri.edu}}\\
Department of Statistics and Data Science, University of Missouri
}

\date{}

\begin{document}
\maketitle

\begin{abstract}
Global water-level systems have undergone large changes in recent years. Moderate Resolution Imaging Spectroradiometer (MODIS) land cover data can be used to monitor and understand these changes. However, such data are extremely high-dimensional making them difficult to process while accounting for all sources of uncertainty. Bayesian hierarchical models are commonly implemented using Markov Chain Monte Carlo (MCMC) methods which can be computationally expensive for large datasets. Exact Posterior Regression (EPR) provides an efficient alternative for inference in latent Gaussian hierarchical models under exponential families without requiring MCMC or approximate Bayesian strategies. EPR has not been developed for multinomial data, and we extend this class of models to the multinomial setting. Additionally, we explore signal-to-noise covariance properties of discrepancy models, which we expect to be present in our data. We extend EPR to multinomial likelihoods via a stick-breaking construction to analyze water-level changes. Through simulation studies, we compare the proposed multinomial EPR approach with alternative models implemented using Gibbs sampling and the Integrated Nested Laplace Approximation (INLA). We evaluate each approach based on regression parameter estimation, category probability estimation, and computational efficiency. Overall, EPR and INLA achieve substantial reductions in computation time while EPR shows improved metrics in estimation and prediction. 

\end{abstract}


\section{Introduction}
\label{sec:intro}
Climate change has affected water-level systems in the Great Lakes region by changing precipitation patterns, evaporation rates, and runoff processes \citep{gronewold2016, kayastha2022}. The Great Lakes, which contain approximately one-fifth of the world's surface fresh water, have experienced substantial water-level fluctuations over the past several decades, with levels varying by more than two meters in response to changes in precipitation, evaporation, and basin runoff \citep{xue2022, kayastha2022}. These fluctuations include a sustained period of below-average water levels from 1998 to 2013, followed by rapid increases that culminated in record-high water levels on several lakes in 2017, 2019, and 2020 \citep{gronewold2016, gronewold2021}. Such variability poses significant risks to ecosystems, commercial shipping, hydropower generation, and coastal communities with high water levels contributing to shoreline erosion and flooding \citep{gronewold2021, kayastha2022}. Although previous studies have investigated Great Lakes water-level variability using hydrological models and historical water-level records, these approaches typically model water level as a continuous variable rather than directly characterizing spatial transitions between land and water cover categories. This motivates an analysis of land-water transitions across the Great Lakes and surrounding regions of Canada and the northern United States from 2010 to 2020 that directly models categorical changes among water-to-water, water-to-land, land-to-water, and land-to-land transitions. To our knowledge, no existing study has modeled large-scale land-water classification transitions using a stick-breaking multinomial spatial Bayesian framework.

Land-water transitions in the Great Lakes region may also involve feedback among receding or rising water, temperature, and other climate-related variables. In particular, changes in precipitation, evaporation, and basin runoff are major drivers of Great Lakes water-level variability \citep{gronewold2016, kayastha2022}. At the same time, changes in ice cover can alter lake temperature, evaporation, and other surface conditions, creating feedbacks within the regional climate system \citep{austin2007,vancleave2014,kayastha2022}. In observational studies such as ours, many relevant climate-related processes, including ice cover, humidity, wind, and atmospheric carbon dioxide, are not directly measured or controlled for. These unobserved processes may introduce dependence between the model's residual error and the latent transition probabilities, which standard spatial models do not account for \citep{bradley2020}. This motivates a modeling framework that explicitly captures such dependence.

\begin{figure}[htbp]
    \centering
    \includegraphics[width=0.9\textwidth]{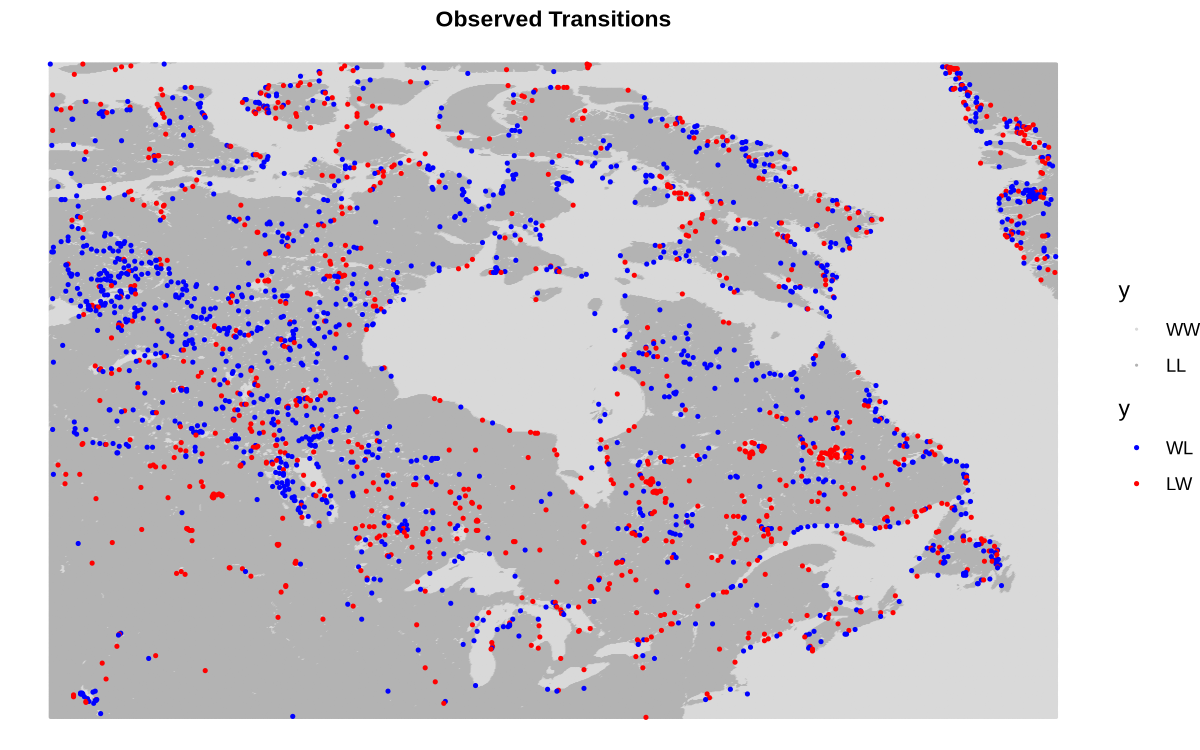}
    \caption{Observed land-water transition locations across the Great Lakes and surrounding regions of Canada and the northern United States. Blue points indicate water-to-land (WL) transitions, and red points indicate land-to-water (LW) transitions.}
    \label{fig:plot_data}
\end{figure}

We consider the Moderate Resolution Imaging Spectroradiometer (MODIS) land cover product (MCD12C1, Collection 6.1), which provides annual global land cover classifications at 0.05-degree spatial resolution \citep{friedl2022}. Using data from 2010 and 2020, we classify each pixel into one of four transition categories: water-to-water (WW), water-to-land (WL), land-to-water (LW), and land-to-land (LL). Our spatial domain covers the Great Lakes and surrounding regions of Canada and the northern United States (longitudes $-115$ to $-50$, latitudes $40$ to $75$). The resulting dataset comprises 910,000 pixel-level observations. We plot the observed land-water transition locations in Figure \ref{fig:plot_data}. The WW and LL categories dominate the dataset, while the transition categories WL and LW are rare events, with only 2,481 observations combined. This class imbalance presents additional challenges for statistical modeling, as the model must accurately distinguish rare transitions from the overwhelmingly common persistence categories. 

We include MODIS land surface temperature (LST)  \citep{WanHookHulley2021} and ERA5-Land precipitation \citep{MunozSabater2019,MunozSabater2021} as covariates for both 2010 and 2020. The MODIS LST product provides surface temperature estimates over both terrestrial surfaces and inland water bodies, allowing us to characterize surface thermal conditions across the Great Lakes region \citep{WanHookHulley2021}. Previous studies have also used MODIS LST to detect changes in land cover \citep{vanLeeuwen2011}. The large spatial scale of the land-water transition dataset, with 910,000 locations, makes traditional Bayesian approaches computationally challenging. Integrated Nested Laplace Approximations \citep[INLA;][]{rue2009} offer a faster alternative to Markov chain Monte Carlo (MCMC) but rely on approximations to posterior marginals. Standard MCMC methods (e.g., Gibbs sampling), while asymptotically exact, can suffer from convergence difficulties and excessive computational costs at this scale. For example, \citet{zhou2026} reported that MCMC failed to converge after 10,000 iterations on a dataset of approximately 100,000 observations, roughly one-ninth the size of our dataset. These limitations motivate the need for a computationally efficient Bayesian framework that can scale to nearly one million observations while maintaining exact posterior inference.

Our first contribution is methodological. We propose a multinomial spatial Bayesian model for land-water transition data using a stick-breaking representation combined with the exact posterior regression (EPR) framework of \citet{bradley2025}. The stick-breaking construction decomposes the four-category multinomial response (water-to-water, water-to-land, land-to-water, and land-to-land) into a sequence of conditional binomial models, each of which can be analyzed within the EPR framework. This approach allows us to directly sample independent replicates from the exact posterior distribution without the need for MCMC, thereby avoiding issues related to convergence diagnostics and burn-in periods. The resulting method scales to datasets with nearly one million spatial locations, which would be computationally challenging  under traditional Bayesian models fitted using MCMC.

Our second contribution addresses the signal-to-noise dependence that may arise from unmeasured climate processes. In observational studies such as ours, climate-related variables including atmospheric carbon dioxide, soil moisture, and wind patterns are not directly measured and may affect the processes underlying land-water transitions. Consequently, the traditional assumption of independence between the latent signal and the additive error term may be restrictive. Following the framework of \citet{bradley2020} and \citet{bradley2025}, we incorporate a discrepancy term that explicitly models the cross-covariance between the signal and the noise. This produces an identifiable covariance structure in the posterior distribution that captures dependence between the latent signal and additive noise. We provide a theoretical result characterizing this cross-covariance structure, extending related signal-to-noise dependence results to the multinomial stick-breaking setting.

Our third contribution is to provide an empirical analysis of large-scale land-water transitions in the Great Lakes region. We apply our model to MODIS land cover data over a spatial domain from the Great Lakes and surrounding regions of Canada and the northern United States, comprising approximately 910,000 pixel-level observations from 2010 and 2020. Through this analysis, we identify spatial patterns in land-water transition probabilities, examine the associations of land surface temperature and precipitation with the transition probabilities, and evaluate predictive performance using the area under the receiver operating characteristic curve (AUC). Our analysis provides spatially resolved information on the geographic patterns of land-water change across the Great Lakes region at nearly one million locations.

The remainder of this paper is organized as follows. In Section \ref{sec:review_gcm_epr}, we briefly describe generalized conjugate multivariate (GCM) distributions and exact posterior regression (EPR). In Section \ref{sec:methods}, we specify our multinomial spatial model and provide theoretical results regarding on the posterior distribution and the signal-to-noise cross-covariance structure. Section \ref{sec:simulation} presents a simulation study in which we compare our method with existing approaches in terms of predictive accuracy and computational efficiency. We apply our model to the MODIS land cover dataset and analyze land-water transitions across the Great Lakes region from 2010 to 2020 in Section \ref{sec:application}. A discussion follows in Section \ref{sec:discussion}.

\section{Preliminaries: Review of GCM and EPR}
\label{sec:review_gcm_epr}
In this section, we introduce the GCM and EPR framework of \citet{bradley2025} that will be used for posterior inference in the proposed multinomial model developed in Section~\ref{sec:methods}. We focus only on the components needed for the multinomial setting considered in this paper.

\subsection{Natural exponential-family form}
Following \citet{bradley2025}, let \(Z\) denote a generic response variable and let \(Y\) denote the corresponding natural parameter. A natural exponential-family density can be written as
\begin{align}
    f_k(Z \mid Y,b_k)
    =
    \exp\{ZY-b_k\psi_k(Y)+c_k(Z)\},
    \qquad
    Z\in\mathcal{Z}_k,\quad Y\in\mathcal{Y}_k,
    \label{eq:nef_clinch}
\end{align}
where \(b_k\) is a known scale, \(\psi_k(\cdot)\) is the unit log-partition function, and \(k\) indexes the exponential-family type.

For multiple independent components, let \(\boldsymbol{Z}=(Z_1,\ldots,Z_M)'\), \(\boldsymbol{Y}=(Y_1,\ldots,Y_M)'\).
For the \(j\)th component, let \(k_j\) denote the corresponding exponential-family type, and let \(b_j\) denote the corresponding known scale. Define $\boldsymbol{b}=(b_1,\ldots,b_M)'$ and $\boldsymbol{\psi}(\boldsymbol{Y}) =
\{\psi_{k_1}(Y_1),\ldots,\psi_{k_M}(Y_M)\}'$.

Then the joint likelihood kernel can be written as
\begin{align}
    f(\boldsymbol{Z}\mid\boldsymbol{Y},\boldsymbol{b})
    \propto
    \exp\left\{
        \boldsymbol{Z}'\boldsymbol{Y}
        -
        \boldsymbol{b}'\boldsymbol{\psi}(\boldsymbol{Y})
    \right\}.
    \label{eq:vector_nef}
\end{align}

The corresponding Diaconis--Ylvisaker conjugate prior for
\(\boldsymbol{Y}\) has the form \citep{diaconis1979}
\begin{align}
    q(\boldsymbol{Y}\mid\boldsymbol{\alpha},\boldsymbol{\kappa})
    \propto
    \exp\left\{
        \boldsymbol{\alpha}'\boldsymbol{Y}
        -
        \boldsymbol{\kappa}'\boldsymbol{\psi}(\boldsymbol{Y})
    \right\},
    \label{eq:dy_kernel}
\end{align}
where \(\boldsymbol{\alpha}\) and \(\boldsymbol{\kappa}\) are hyperparameters. The posterior remains in the same conjugate family because the likelihood in \eqref{eq:vector_nef} and the prior in \eqref{eq:dy_kernel} have the same functional form.

\subsection{Generalized conjugate multivariate distribution}
The GCM distribution extends the multivariate conjugate construction by introducing a location vector, an invertible dependence matrix, and a scaling matrix that may depend on hyperparameters. Following \citet{bradley2025}, let $\boldsymbol{w} \in \mathbb{R}^{M}$ denote a vector whose components are independent Diaconis--Ylvisaker random variables with parameters $\boldsymbol{\alpha}$, $\boldsymbol{\kappa}$, and unit log-partition functions collected in $\boldsymbol{\psi}$. The GCM random vector is defined through the transformation
$\boldsymbol{y}=\boldsymbol{\mu}+\mathbf{V}\mathbf{D}(\theta)\boldsymbol{w}$, where $\boldsymbol{\mu}$ is a location vector, $\mathbf{V}$ is an invertible dependence matrix, \(\mathbf{D}(\theta)\) is a nonsingular scaling matrix depending on hyperparameters \(\theta\), and \(\theta\sim\pi(\theta)\).

Equivalently, the vector $\boldsymbol{w}$ can be written as 
\[\boldsymbol{w}=
\mathbf{D}(\theta)^{-1}\mathbf{V}^{-1}(\boldsymbol{y}-\boldsymbol{\mu}).\]
Therefore, after marginalizing over \(\theta\), the density of
\(\boldsymbol{y}\) has kernel
\[
    f(\boldsymbol{y}
    \mid
    \boldsymbol{\mu},\mathbf{V},
    \boldsymbol{\alpha},\boldsymbol{\kappa})
    \propto
    \int
    \pi(\theta)
    \left|\mathbf{D}(\theta)^{-1}\mathbf{V}^{-1}\right|
    \exp\left[
        \boldsymbol{\alpha}'
        \mathbf{D}(\theta)^{-1}\mathbf{V}^{-1}
        (\boldsymbol{y}-\boldsymbol{\mu})
        -
        \boldsymbol{\kappa}'
        \boldsymbol{\psi}
        \left\{
        \mathbf{D}(\theta)^{-1}\mathbf{V}^{-1}
        (\boldsymbol{y}-\boldsymbol{\mu})
        \right\}
    \right]
    d\theta .
\]
We write this distribution as
$\boldsymbol{y}\sim GCM(\boldsymbol{\alpha}, \boldsymbol{\kappa}, \boldsymbol{\mu}, \mathbf{V}, \pi, \mathbf{D}; \boldsymbol{\psi})$. The posterior distribution of our proposed model will follow a specific GCM. A conditional GCM (cGCM) distribution is obtained by conditioning a subset of a GCM random vector on the remaining components \citep{bradley2025}. In Section~\ref{sec:process_model}, we use a cGCM prior for the fine-scale component of the proposed multinomial model.

\section{Model}
\label{sec:methods}
In this section, we specify the statistical model used for analyzing multinomial land-water transition data. Section~\ref{sec:data_model} introduces the data model, including the multinomial likelihood and its stick-breaking construction. Section~\ref{sec:process_model} provides the latent process model with a discrepancy term. Section~\ref{sec:post_gcm} provides a posterior GCM representation, and Section~\ref{sec:signal-to-noise} examines the resulting signal-to-noise dependence.

\subsection{Data Model}\label{sec:data_model}
Let $\{\boldsymbol{s}_i: i = 1, \ldots, n\}$ denote a collection of spatial locations. At each location $\boldsymbol{s}_i$, let $Z_{\ell}(\boldsymbol{s}_i)\in\{1,\ldots,K\}$ denote a categorical response, where $K$ is the number of categories and $\ell = 1,\ldots, L$ represents repeated measurements at site $\boldsymbol{s}_i$. Let $ \boldsymbol{\pi}(\boldsymbol{s}_i) = \{\pi_1(\boldsymbol{s}_i),\ldots,\pi_K(\boldsymbol{s}_i)\}'$ denote the corresponding vector of category probabilities, where $\pi_k(\boldsymbol{s}_i)=\Pr\{Z_{\ell}(\boldsymbol{s}_i)=k\}$ and $\sum_{k=1}^K \pi_k(\boldsymbol{s}_i)=1$. Conditional on $\boldsymbol{\pi}(\boldsymbol{s}_i)$, we assume that
\[
Z_{\ell}(\boldsymbol{s}_i) \mid \boldsymbol{\pi}(\boldsymbol{s}_i) \sim \mathrm{Categorical}\{\boldsymbol{\pi}(\boldsymbol{s}_i)\},
\qquad
\ell=1,\ldots,L, \quad i = 1, \ldots, n
\]

\noindent Define the category counts
\[
Y_{ki}
=
\sum_{\ell=1}^{L}
\mathbf{1}\{Z_{\ell}(\boldsymbol{s}_i)=k\},
\qquad k=1,\ldots,K,
\]
where $\mathbf{1}\{Z_{\ell}(\boldsymbol{s}_i)=k\}= 1$  if $Z_{\ell}(\boldsymbol{s}_i)=k$ and is equal to zero otherwise. Let $\boldsymbol{Y}_i = (Y_{1i},\ldots,Y_{Ki})'.$ Then $Y_{ki}$ denotes the number of repeated measurements at location $\boldsymbol{s}_i$ that belong to category $k$, and $m_i = \sum_{k=1}^K Y_{ki}=L$ denotes the total count at location $\boldsymbol{s}_i$. Hence,
\[
\boldsymbol{Y}_i
\mid
\boldsymbol{\pi}(\boldsymbol{s}_i)
\sim
\mathrm{Multinomial}\{m_i,\boldsymbol{\pi}(\boldsymbol{s}_i)\}.
\]

\noindent In our application, $L=1$ and the spatial locations correspond to pixels on a regular grid over the Great Lakes region of North America. At each location, we observe land cover classifications from the Moderate Resolution Imaging Spectroradiometer (MODIS) Land Cover Type product (MCD12C1) at two time points, 2010 and 2020 \citep{friedl2022}. These two classifications are used to construct one categorical land-water transition response at each pixel. Therefore, $m_i=1$ in our application.

\begin{align}\label{eq:categories}
Z_1(\boldsymbol{s}_i) = \begin{cases}
1, & \text{if classified as water in 2010 and water in 2020 (WW)}, \\
2, & \text{if classified as water in 2010 and land in 2020 (WL)}, \\
3, & \text{if classified as land in 2010 and water in 2020 (LW)}, \\
4, & \text{if classified as land in 2010 and land in 2020 (LL)},
\end{cases}
\end{align}

We use a stick-breaking construction to rewrite the multinomial likelihood as a product of binomial likelihood factors \citep[e.g.,][]{linderman2015, bradley2019}. The idea is to assign categories sequentially. At the first step, we consider whether the response belongs to category 1. If it does not, then at the second step we consider whether it belongs to category 2 among the remaining categories. This process continues through category $K-1$, while category $K$ is determined by the remaining probability. This representation is useful because it gives the binomial exponential-family form needed for the conjugate framework described in Section~\ref{sec:post_gcm}.

For $k=1,\ldots,K-1$, define the remaining trial count at location $\boldsymbol{s}_i$ as $n_{ki} = m_i-\sum_{j<k}Y_{ji}.$ The $n_{ki}$ is the number of observations at location $\boldsymbol{s}_i$ that have not already been assigned to categories $1,\ldots,k-1$. Thus, at the $k$-th stick-breaking step, $n_{ki}$ is the number of observations still available to be assigned to category $k$.

We also define the conditional probability
\[
p_{ki}
=
\Pr\{Z_{\ell}(\boldsymbol{s}_i)=k
\mid
Z_{\ell}(\boldsymbol{s}_i)\notin\{1,\ldots,k-1\}\}.
\]
Equivalently, in terms of the original multinomial probabilities,
\[
p_{ki}
=
\frac{\pi_k(\boldsymbol{s}_i)}
{1-\sum_{j<k}\pi_j(\boldsymbol{s}_i)}.
\]
Thus, $p_{ki}$ is the conditional probability that an observation belongs to category $k$, given that it does not belong to any of the previous categories $1,\ldots, k-1$. For notational simplicity, write $\pi_{ki}=\pi_k(\boldsymbol{s}_i)$. The
multinomial probability mass function can be written as
\begin{align}
f\left(\boldsymbol{Y}_i\mid \boldsymbol{\pi}(\boldsymbol{s}_i)\right)
&=
\frac{m_i!}{\prod_{k=1}^{K}Y_{ki}!}
\prod_{k=1}^{K}
\pi_{ki}^{Y_{ki}} \notag \\
&=
\prod_{k=1}^{K-1}
\binom{n_{ki}}{Y_{ki}}
p_{ki}^{Y_{ki}}
(1-p_{ki})^{n_{ki}-Y_{ki}},
\label{eq:product_binomial}
\end{align}
where the second equality follows from the stick-breaking
reparameterization with
\[
n_{ki}
=
m_i-\sum_{j<k}Y_{ji},
\qquad
p_{ki}
=
\frac{\pi_{ki}}
{1-\sum_{j<k}\pi_{ji}},
\qquad k=1,\ldots,K-1.
\]

Assuming a logit link function on $p_{ki}$, that is, $\nu_{ki} \equiv \text{logit}(p_{ki})
= \log\left\{\frac{p_{ki}}{1-p_{ki}}\right\}$ allows us to rewrite the product of binomials in \eqref{eq:product_binomial} as
\begin{align}\label{eq:logit_likelihood}
f_{\text{MN}}\left(\boldsymbol{Y}_i\mid \{\nu_{ki}\}_{k=1}^{K-1}\right)
=
\prod_{k=1}^{K-1}
\binom{n_{ki}}{Y_{ki}}
\frac{\exp(Y_{ki}\nu_{ki})}
{\{1+\exp(\nu_{ki})\}^{n_{ki}}}.
\end{align}
This formulation shows that, conditional on $n_{ki}>0$, each component of the stick-breaking likelihood has a natural exponential-family form with natural parameter $\nu_{ki}$, known scale $n_{ki}$, and unit log-partition function $\log\{1+\exp(\nu_{ki})\}$.

Let $\boldsymbol{\nu} = (\nu_{ki}: k=1,\ldots,K-1,\ i=1,\ldots,n,\ n_{ki}>0)'$
denote the vector of logit parameters. Similarly, let $\mathbf{y} = (Y_{ki}: k=1,\ldots,K-1,\ i=1,\ldots,n,\ n_{ki}>0)'$ denote the corresponding response vector, and let $\mathbf{n} =
(n_{ki}: k=1,\ldots,K-1,\ i=1,\ldots,n,\ n_{ki}>0)'$ denote the vector of binomial totals. Then the full likelihood can be written as
\begin{align}\label{eq:full_likelihood}
f(\mathbf{y}\mid \boldsymbol{\nu})
\propto
\exp\left[
\mathbf{y}'\boldsymbol{\nu}
-
\mathbf{n}'\log\{\mathbf{1}+\exp(\boldsymbol{\nu})\}
\right],
\end{align}
\noindent which has the natural exponential-family form required for the GCM/EPR construction described in Section~\ref{sec:post_gcm} \citep{bradley2025, zhou2026}

\subsection{Process Model}
\label{sec:process_model}
We specify a spatial mixed effects model for the latent logit parameters introduced in Section~\ref{sec:data_model}. Recall that $\nu_{ki}=\mathrm{logit}(p_{ki})$, where $p_{ki}$ is the conditional stick-breaking probability for category $k$ at location $\boldsymbol{s}_i$. Following standard hierarchical spatial modeling ideas \citep{cressie2011} and the discrepancy and EPR constructions developed in \citet{bradley2020}, \citet{bradley2025}, and \citet{zhou2026}, we represent the latent logit process using large-scale, small-scale, fine-scale variation, together with a discrepancy term. Specifically, for $k=1,\ldots,K-1$ and locations $\boldsymbol{s}_i$ with $n_{ki}>0$, we assume
\begin{align}
\nu_{ki}
=
\boldsymbol{x}_{ki}'\boldsymbol{\beta}
+
\boldsymbol{g}_{ki}'\boldsymbol{\eta}
+
\xi_{ki}
-
\delta_{\nu,ki}.
\label{eq:process_model}
\end{align}
Here, $\boldsymbol{x}_{ki}$ and $\boldsymbol{g}_{ki}$ denote the rows of the stacked covariate and basis-function matrices corresponding to location $\boldsymbol{s}_i$ and stick-breaking level $k$. These rows are constructed by repeating the location-level covariates and basis functions across the $K-1$ stick-breaking levels and then retaining only rows with $n_{ki}>0$.

The term $\boldsymbol{x}_{ki}'\boldsymbol{\beta}$ represents large-scale variation in the conditional log-odds of the stick-breaking probability. The vector $\boldsymbol{x}_{ki}$ contains observed covariates for location $\boldsymbol{s}_i$, and $\boldsymbol{\beta}$ is the corresponding vector of regression coefficients. We assume
\[
\boldsymbol{\beta}\mid \sigma_\beta^2
\sim
N(\boldsymbol{\delta}_\beta,\sigma_\beta^2\mathbf{I}_p),
\qquad
\sigma_\beta^2\sim IG(a_\beta,b_\beta),
\]
where $\boldsymbol{\delta}_\beta$ is the discrepancy term associated with $\boldsymbol{\beta}$, defined in Section~\ref{sec:post_gcm}, and $\sigma_\beta^2$ controls the prior variability of the fixed effects. In our application, the covariates include an intercept, land surface temperature and precipitation measurements from 2010 and 2020, and missing indicators for these climate variables. This component captures broad relationships between observed climate conditions and the conditional probabilities of land-water transition categories.

The term $\boldsymbol{g}_{ki}'\boldsymbol{\eta}$ represents small-scale spatial variability that remains after accounting for the observed covariates. The vector $\boldsymbol{g}_{ki}$ contains spatial basis functions evaluated at location $\boldsymbol{s}_i$, and $\boldsymbol{\eta}$ is the corresponding vector of spatial random effects. We assume
\[
\boldsymbol{\eta}\mid \sigma_\eta^2
\sim
N(\boldsymbol{\delta}_\eta,\sigma_\eta^2\mathbf{I}_r),
\qquad
\sigma_\eta^2\sim IG(a_\eta,b_\eta),
\]
where $\boldsymbol{\delta}_\eta$ is the discrepancy term associated with $\boldsymbol{\eta}$, defined in Section~\ref{sec:post_gcm}, and $\sigma_\eta^2$ controls the prior variability of the spatial random effects. In our analysis, we use Gaussian radial basis functions. Let $\{\boldsymbol{c}_1,\ldots,\boldsymbol{c}_r\}$ denote the knot locations. The $j$th basis function evaluated at location $\boldsymbol{s}_i$ is
\begin{align}
G_{ij}
=
\exp\left\{
-b\|\boldsymbol{s}_i-\boldsymbol{c}_j\|^2
\right\},
\qquad i=1,\ldots,n,\quad j=1,\ldots,r,
\label{eq:basis}
\end{align}
where $b>0$ is a bandwidth parameter. In addition, the knot set is formed by combining a regular grid of knots with additional knots near observed land-water transition locations. Details on the number of knots and bandwidth choice are provided in Section~\ref{sec:application}.

The term $\xi_{ki}$ represents fine-scale variability in the conditional log-odds. In the land-water transition application, this component captures local pixel-level heterogeneity and other fine-scale variation that is not represented by the observed covariates or the basis expansion.

Let $N = \sum_{i=1}^{n}\sum_{k=1}^{K-1}\mathbf{1}(n_{ki}>0)$ denote the number of nondegenerate stick-breaking likelihood contributions, and let $\boldsymbol{\nu}$ and $\boldsymbol{\xi}$ denote the corresponding stacked vectors of $\nu_{ki}$ and $\xi_{ki}$, respectively. In a traditional latent Gaussian process model, the stacked fine-scale term is often assumed to be Gaussian. Following the EPR framework of \citet{bradley2025}, we instead assign \(\boldsymbol{\xi}\) a conditional GCM (cGCM) prior. This specification reduces to a Gaussian fine-scale model when $\alpha_\xi=0$, whereas we take $\alpha_\xi>0$ for the cGCM specification. The specific cGCM prior used for $\boldsymbol{\xi}$ is given by
\begin{align}
f(\boldsymbol{\xi}\mid
\boldsymbol{\beta},\boldsymbol{\eta},\theta,\mathbf{q})
&\propto
\exp\Bigg[
\alpha_\xi\mathbf{1}_N'\boldsymbol{\nu}
-
2\alpha_\xi\mathbf{1}_N'
\boldsymbol{\psi}_3(\boldsymbol{\nu})
-
\frac{1}{2}
\left(
\sigma_\xi^{-1}\boldsymbol{\xi}
-
\boldsymbol{\delta}_\xi
\right)'
\left(
\sigma_\xi^{-1}\boldsymbol{\xi}
-
\boldsymbol{\delta}_\xi
\right)
\Bigg],
\label{eq:cgcm_xi}
\end{align}
Here, $\sigma_\xi^2$ and $\alpha_\xi$ are fixed.
The discrepancy components $\boldsymbol{\delta}_{\nu}$ and
$\boldsymbol{\delta}_{\xi}$ are defined in Section~\ref{sec:post_gcm}.

Finally, \(\delta_{\nu,ki}\) denotes the discrepancy term associated with the latent logit process. This term is introduced to allow for dependence between the latent signal and the residual error, rather than imposing the classical assumption that they are independent. In observational environmental studies, relevant processes may be unmeasured. In our application, unmeasured processes such as ice cover, wind, humidity, soil moisture, and atmospheric carbon dioxide may affect both the observed climate covariates and the land-water transition probabilities. Thus, the discrepancy term provides a mechanism for allowing dependence between the latent transition process and residual variation.

After stacking over all locations and stick-breaking levels with \(n_{ki}>0\),
the process model can be written as
\begin{align}
\boldsymbol{\nu}
=
\mathbf{X}\boldsymbol{\beta}
+
\mathbf{G}\boldsymbol{\eta}
+
\boldsymbol{\xi}
-
\boldsymbol{\delta}_{\nu},
\label{eq:stacked_process_model}
\end{align}
where \(\boldsymbol{\nu}\) is the vector of stick-breaking logits,
\(\mathbf{X}\) is the stacked covariate matrix, \(\mathbf{G}\) is the stacked
basis-function matrix, \(\boldsymbol{\xi}\) is the fine-scale variability term,
and \(\boldsymbol{\delta}_{\nu}\) is the discrepancy term associated with the latent logit process.

\subsection{Posterior GCM Representation}
\label{sec:post_gcm}

We now derive the posterior representation implied by the stick-breaking likelihood in Section~\ref{sec:data_model} and the process model in Section~\ref{sec:process_model}. By the stick-breaking construction, the multinomial likelihood is written as a product of conditional binomial likelihood factors. With the logit transformation, $\nu_{ki} = \log\left\{ \frac{p_{ki}}{1-p_{ki}} \right\},$ each stick-breaking component has natural exponential-family form. Therefore, the stacked likelihood can be embedded in the generalized conjugate multivariate (GCM) framework.

Recall that $N = \sum_{i=1}^{n} \sum_{k=1}^{K-1} \mathbf{1}(n_{ki}>0)$ denotes the number of nondegenerate stick-breaking likelihood contributions. Let $\boldsymbol{\zeta} = (\boldsymbol{\xi}',\boldsymbol{\beta}',\boldsymbol{\eta}')'$ be the stacked vector of unknown model effects. Define the matrix  \[ \mathbf{H} = \begin{pmatrix} \mathbf{I}_{N} & \mathbf{X} & \mathbf{G}\\ \mathbf{0}_{p,N} & \mathbf{I}_{p} & \mathbf{0}_{p,r}\\ \mathbf{0}_{r,N} & \mathbf{0}_{r,p} & \mathbf{I}_{r}\\ \mathbf{I}_{N} & \mathbf{0}_{N,p} & \mathbf{0}_{N,r} \end{pmatrix}. \]  
Let $M=2N+p+r$, so that $\mathbf{H}\in\mathbb{R}^{M\times(N+p+r)}$, and  
\(\mathbf{Q}\in\mathbb{R}^{M\times N}\) be a matrix whose columns form an orthonormal basis for the orthogonal complement of the column space of \(\mathbf{H}\), so that $\mathbf{H}'\mathbf{Q}=\mathbf{0}$, $\mathbf{Q}'\mathbf{Q}=\mathbf{I}_{N}$,
and $\mathbf{Q}\mathbf{Q}' = \mathbf{I}_{M} - \mathbf{H}(\mathbf{H}'\mathbf{H})^{-1}\mathbf{H}'$. Following \citet{bradley2025}, the discrepancy is introduced through this orthogonal complement by letting $\boldsymbol{\delta} = (\boldsymbol{\delta}_{\nu}', \boldsymbol{\delta}_{\beta}', \boldsymbol{\delta}_{\eta}', \boldsymbol{\delta}_{\xi}')' = - \mathbf{D}(\theta)^{-1}\mathbf{Q}\mathbf{q}$, where $\mathbf{D}(\theta)^{-1} = \mathrm{blkdiag} \left( \mathbf{I}_{N}, D_{\beta}(\theta)^{-1}, D_{\eta}(\theta)^{-1}, \sigma_{\xi}^{-1}\mathbf{I}_{N} \right)$, and $\mathbf{q}\in\mathbb{R}^{N}$ is a free discrepancy parameter.

\begin{proposition}[Posterior GCM representation]
\label{prop:gcm_epr}
Assume the stick-breaking binomial likelihood in
Section~\ref{sec:data_model}, the process model in
Section~\ref{sec:process_model}, and the prior specifications
for $\boldsymbol{\beta}$, $\boldsymbol{\eta}$, and $\boldsymbol{\xi}$.
Let $\boldsymbol{\zeta}=(\boldsymbol{\xi}',\boldsymbol{\beta}',\boldsymbol{\eta}')'$,
and let $\mathbf{H}$ and $\mathbf{Q}$ be as defined in
Section~\ref{sec:post_gcm}. Then
\[
  (\boldsymbol{\zeta}',\mathbf{q}')'\mid\mathbf{y}
  \sim
  \mathrm{GCM}\!\left(
    \boldsymbol{\alpha}_{M},\,
    \boldsymbol{\kappa}_{M},\,
    \mathbf{0}_{M,1},\,
    \mathbf{V}_{M},\,
    \pi,\,
    \mathbf{D};\,
    \boldsymbol{\psi}_{M}
  \right),
\]
where $M=2N+p+r$, $\mathbf{V}_{M}^{-1}=(\mathbf{H},\mathbf{Q})$,
$\mathbf{D}(\theta)^{-1}=\mathrm{blkdiag}(\mathbf{I}_{N},\mathbf{D}_{\beta}(\theta)^{-1},
\mathbf{D}_{\eta}(\theta)^{-1},\sigma_{\xi}^{-1}\mathbf{I}_{N})$,
the shape and scale vectors are
\[
  \boldsymbol{\alpha}_{M}
  =\bigl(\mathbf{y}'+\alpha_{\xi}\mathbf{1}_{N}',\;
         \mathbf{0}_{1,N+p+r}\bigr)',
  \qquad
  \boldsymbol{\kappa}_{M}
  =\bigl(\mathbf{n}'+2\alpha_{\xi}\mathbf{1}_{N}',\;
         \tfrac{1}{2}\mathbf{1}_{1,N+p+r}\bigr)',
\]
and the unit log-partition function is
$\boldsymbol{\psi}_{M}(h)
 =\bigl(\psi_{3}(h_{1}),\ldots,\psi_{3}(h_{N}),
         \psi_{1}(h_{1}^{*}),\ldots,\psi_{1}(h_{N+p+r}^{*})\bigr)'$
for any $(2N+p+r)$-dimensional real-valued vector
$h=(h_{1},\ldots,h_{N},h_{1}^{*},\ldots,h_{N+p+r}^{*})'$.

Equivalently, the posterior admits the EPR representation 
\[
\begin{pmatrix}
\boldsymbol{\zeta}\\
\mathbf{q}
\end{pmatrix}
=
\mathbf{V}_{M}\mathbf{D}(\theta)\boldsymbol{w}_{M}
=
\begin{pmatrix}
(\mathbf{H}'\mathbf{H})^{-1}\mathbf{H}'\\
\mathbf{Q}'
\end{pmatrix}
\mathbf{D}(\theta)\boldsymbol{w}_{M},
\]
where the components of $\boldsymbol{w}_{M}$ are independent
Diaconis--Ylvisaker random variables with parameters
$\boldsymbol{\alpha}_{M}$ and $\boldsymbol{\kappa}_{M}$.
\end{proposition}

The proof of Proposition~\ref{prop:gcm_epr} is provided in Appendix~\ref{app:proof_gcm_epr}. One can sample from this GCM directly, using replicates Proposition~\ref{prop:gcm_epr} the regression form presented in as a result. In previous papers, these replicates are referred to as "Exact Posterior Regression"  (\cite{bradley2025}; \citet{clinch2026}; \cite{zhou2026}). Moreover, the blockwise structure defining $\mathbf{H}$ leads to computational speed-ups and only needs to be computed once when simulating multiple replicates.

\subsection{Signal-to-Noise Dependence}
\label{sec:signal-to-noise}
Traditionally, additive residual terms in spatial hierarchical
models are often assumed to be independent of the latent signal. In the present
model, the latent stick-breaking signal is
$
    \boldsymbol{\nu}^{*}
    =
    \mathbf{X}\boldsymbol{\beta}
    +
    \mathbf{G}\boldsymbol{\eta}
    +
    \boldsymbol{\xi},
$
and the stick-breaking linear predictor is written as
$
    \boldsymbol{\nu}
    =
    \boldsymbol{\nu}^{*}
    -
    \boldsymbol{\delta}_{\nu}.
$
Thus, the additive noise contribution entering the linear predictor is
\(-\boldsymbol{\delta}_{\nu}\). In many observational environmental studies,
however, the discrepancy term need not be independent of the
latent signal. For example, unmeasured processes such as ice cover, wind,
humidity, soil moisture, runoff, evaporation, and atmospheric carbon dioxide
may influence both the observed climate covariates and the land-water
transition probabilities, but are not explicitly included in
\(\boldsymbol{\nu}^{*}\) through the covariates or the spatial basis
representation.

The GCM construction explicitly permits this type of signal-to-noise dependence. In particular, the latent signal and the discrepancy component are represented through complementary projection spaces. The signal is associated with the column space of $\mathbf{H}$, while the discrepancy is associated with its orthogonal complement. As a result, the implied posterior distribution induces a parametric form for the cross signal-to-noise covariance. The following proposition gives the corresponding posterior cross-covariance for the proposed model.

\begin{proposition}[Signal-to-noise cross-covariance]
\label{prop:signal_noise}
Under the GCM posterior representation of
Proposition~\ref{prop:gcm_epr}, let
$\boldsymbol{\nu}^{*}=\mathbf{X}\boldsymbol{\beta}
 +\mathbf{G}\boldsymbol{\eta}+\boldsymbol{\xi}$
denote the stick-breaking signal and
$-\boldsymbol{\delta}_{\nu}$ the additive noise contribution
to the natural parameter. Define
$\mathbf{P}_{H}=\mathbf{H}(\mathbf{H}'\mathbf{H})^{-1}\mathbf{H}'$
and $\mathbf{J}_{\nu}=(\mathbf{I}_{N},\mathbf{0}_{N,M-N})$.

\noindent\emph{(i) Conditional on $\mathbf{y}$ and $\theta$:}
\[
  \operatorname{Cov}(\boldsymbol{\nu}^{*},-\boldsymbol{\delta}_{\nu}
    \mid\mathbf{y},\theta)
  =\mathbf{J}_{\nu}
   \mathbf{P}_{H}
   \mathbf{D}(\theta)\,
   \operatorname{Cov}(\boldsymbol{w}_{M}\mid\boldsymbol{\alpha}_{M},
     \boldsymbol{\kappa}_{M})\,
   \mathbf{D}(\theta)'
   \bigl\{\mathbf{I}_{M}-\mathbf{P}_{H}\bigr\}
   \mathbf{J}_{\nu}'.
\]

\noindent\emph{(ii) Marginally over $\mathbf{y}$, with $\theta$ fixed:}
\[
  \operatorname{Cov}(\boldsymbol{\nu}^{*},-\boldsymbol{\delta}_{\nu}
    \mid\theta)
  =\mathbf{J}_{\nu}
  \mathbf{P}_{H}
   \mathbf{D}(\theta)
   \Bigl\{
     E_{\mathbf{y}}\!\bigl[\boldsymbol{\Sigma}_{w}(\mathbf{y})\bigr]
     +\operatorname{Cov}_{\mathbf{y}}\!\bigl(
       \boldsymbol{\mu}_{w}(\mathbf{y})\bigr)
   \Bigr\}
   \mathbf{D}(\theta)'
   \bigl\{\mathbf{I}_{M}-\mathbf{P}_{H}\bigr\}
   \mathbf{J}_{\nu}',
\]
where
$\boldsymbol{\mu}_{w}(\mathbf{y}):=E[\boldsymbol{w}_{M}\mid\mathbf{y}]$,
$\boldsymbol{\Sigma}_{w}(\mathbf{y}):=
 \operatorname{Cov}(\boldsymbol{w}_{M}\mid\mathbf{y})$.
\end{proposition}

Since neither term in the braces in Proposition~\ref{prop:signal_noise} (ii) is required to vanish, the marginal cross-covariance between the signal and discrepancy need not be zero. The proof of Proposition~\ref{prop:signal_noise} is provided in Appendix~\ref{app:proof_signal_noise}. This construction is related to the signal-to-noise covariance structure of \citet{zhou2026}, but is developed here for the stick-breaking multinomial likelihood and the stacked logit process $\boldsymbol{\nu}$. In contrast to the posterior cross-covariance conditional on the observed data considered by \citet{zhou2026}, Proposition~\ref{prop:signal_noise} characterizes the marginal cross-covariance between the latent signal and discrepancy. This is useful because the marginal covariance accounts for both the conditional covariance and variation in the conditional means across $\mathbf{y}$.

\section{Simulation Study}
\label{sec:simulation}
In this section, we evaluate the proposed EPR approach against two computational strategies such as a Markov chain Monte Carlo (MCMC) referred to as the Gibbs sampler for the multinomial logit-beta model \citep{bradley2019} and the integrated nested Laplace approximation (INLA) \citep{rue2009}. These computational strategies are applied to different model specifications. In particular, the model fitted with INLA does not include a discrepancy term, whereas the proposed discrepancy model does. Thus, the comparison with INLA also allows us to examine the effect of including a discrepancy term on prediction and regression-effect estimation. Additionally, the discrepancy assumed in \citet{bradley2019}  is different from ours. As such, comparison to \citet{bradley2019} implemented with a Gibbs sampler allows us to compare to alternate discrepancy models. Our goals are to assess (i) the predictive accuracy of the estimated category probabilities at both observed and held-out locations, (ii) the accuracy and uncertainty quantification of the estimates, and (iii) the computational cost of each method.

\subsection{Simulation Design}
\label{sec:sim_design}
We simulate categorical data with $K=5$ categories at $n_{\mathrm{loc}}=50$ spatial locations using the stick-breaking representation of Section~\ref{sec:data_model}. The $N=200$ stacked logits are generated from the linear predictor $\boldsymbol{\nu}=\mathbf{X}\boldsymbol{\beta}+\boldsymbol{G}\boldsymbol{\eta}+\boldsymbol{\xi}$,
where the design matrix $\mathbf{X}=[\mathbf{1},\,\sin(\pi\mathbf{s})]$ is defined on the stacked scale and $\boldsymbol{\beta}=(0.01,-2)'$. The columns of $\boldsymbol{G}$ are the leading $r=100$ Moran's I basis functions \citep{hughes2013moran}, obtained from the eigenvectors of the orthogonal-complement projection $\mathbf{I}_{N}-\mathbf{Q}\mathbf{Q}'$, where $\mathbf{Q}$ spans the column space of $\mathbf{X}$. The random effects $\boldsymbol{\eta}$ and fine-scale term $\boldsymbol{\xi}$ are generated as mean-zero Gaussian vectors, with variances set to one half of the empirical variance of $\mathbf{X}\boldsymbol{\beta}$ and $\boldsymbol{G}\boldsymbol{\eta}$, respectively.

Category probabilities $\pi_{1},\ldots,\pi_{5}$ are obtained from $\boldsymbol{\nu}$ through the stick-breaking transformation. These probabilities, together with location-specific sample sizes $m_{i}\sim\mathrm{Poisson}(50)$, are generated once and held fixed throughout the study.

Conditional on these fixed category probabilities and sample sizes, we draw $50$ independent replicates of the multinomial counts at each location. In each replicates, $10\%$ of the locations is selected uniformly at random and treated as missing. Specifically, when a location is selected, all $K-1$ stick-breaking responses associated with that location are withheld.

For each replicate, EPR draws $B=1{,}000$ independent posterior samples, while the model implemented with Gibbs sampler is run for $B=2{,}000$ iterations with the first $1{,}000$ iterations discarded as burn-in. INLA is fit to the stacked binomial likelihood using a binomial family with a logit link. Its fixed-effect specification includes stick-breaking-step indicators, the covariate, and the Moran's I basis functions. Posterior inference for $\boldsymbol{\pi}$ is based on posterior samples when using Gibbs or EPR, and for INLA we draw $1{,}000$ samples from the fitted approximation using \texttt{inla.posterior.sample}. The stick-breaking transformation is applied to each sample, so that posterior means and pointwise $95\%$ credible intervals for $\boldsymbol{\pi}$ are computed identically across strategies. For each strategy, computation time is recorded as elapsed wall-clock time, with INLA run on a single computational thread. Predictive accuracy for $\boldsymbol{\pi}$ is evaluated separately at the held-out (missing) locations and at the observed locations.

\subsection{Illustration}
\label{sec:sim_illustration}
\begin{figure}[ht]
\centering
\includegraphics[width=\textwidth]{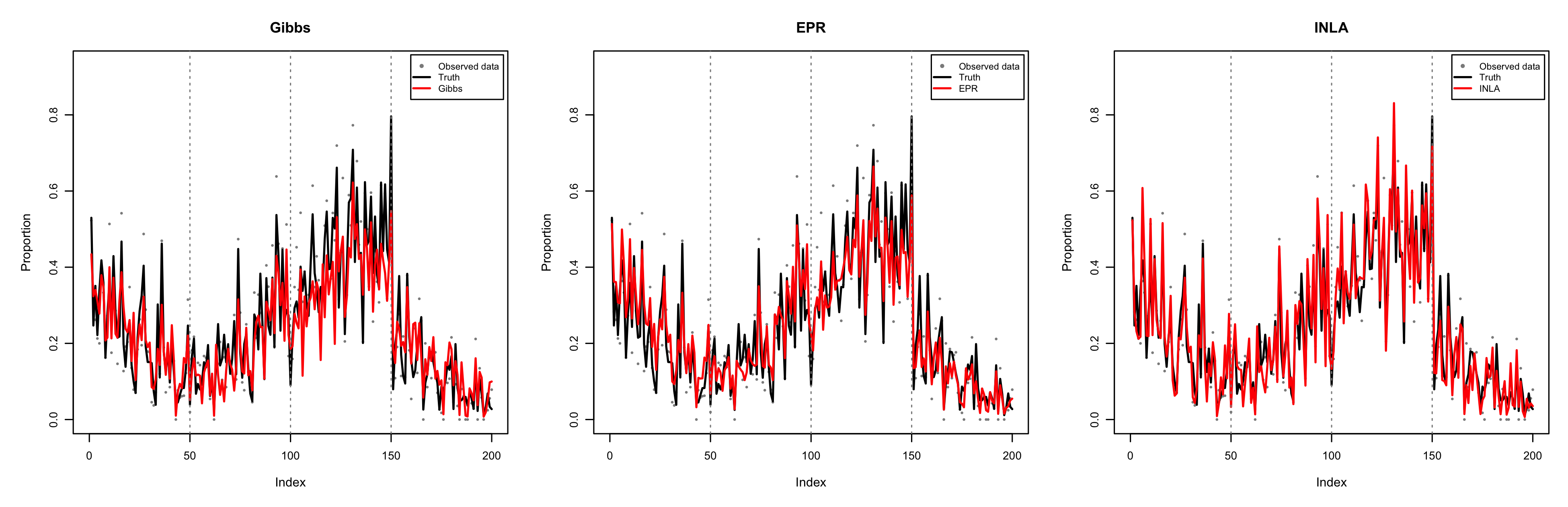}
\caption{Estimated category probabilities for one simulation replicate (replicate 35). In each panel, grey points are the observed empirical proportions, the black line is the true category probability, and the colored line is the posterior mean estimate from Gibbs, EPR, and INLA, respectively. The horizontal axis is the stacked index over the four stick-breaking categories, with dotted vertical lines marking the category boundaries at indices $50$, $100$, and $150$. Estimates close to the true probabilities and the observed proportions indicate accurate recovery of the category probabilities.}
\label{fig:sim_illustration}
\end{figure}

Figure~\ref{fig:sim_illustration} illustrates the estimated category probabilities for one simulation replicate. Each panel plots the first four category probabilities across the stacked index ($1$--$200$), where the observed empirical proportions (grey points) and the true probabilities (black line) are compared with the posterior mean estimates from three different models implemented with Gibbs, EPR, and INLA, respectively. All three methods closely track the true probabilities across the four stick-breaking categories, which are separated by the dotted vertical lines at indices $50$, $100$, and $150$. In this replicate, results using EPR and INLA attain nearly identical prediction error, slightly smaller than that of the Gibbs sampler, while EPR additionally avoids burn-in and Markov chain convergence. This comparison suggests that the independent posterior samples generated by EPR provide predictive inference comparable to models implemented with efficient MCMC and INLA.

\subsection{Multinomial Count Case}
\label{sec:sim_multi_results}
Table~\ref{tab:sim_main} reports the average performance over the $50$ simulated replicates under $10\%$ location-wise missingness. We summarize the mean squared error (MSE) of the estimated first $K-1$ category probabilities at the held-out and observed locations separately, together with the mean elapsed computation time for each strategy.

\begin{table}[!htbp]
\centering
\vspace{0.4em} \small
\begin{tabular}{lccccc}
\hline
 & \multicolumn{2}{c}{Held-out locations} & \multicolumn{2}{c}{Observed locations} & \\
\cmidrule(lr){2-3}\cmidrule(lr){4-5}
Strategy & $\mathrm{MSE}_{\pi}$ mean & $\mathrm{MSE}_{\pi}$ sd & $\mathrm{MSE}_{\pi}$ mean & $\mathrm{MSE}_{\pi}$ sd & CPU (s) \\
\hline
Gibbs & 0.0266 & 0.0177 & 0.0050 & 0.0007 & 5.1551 \\
EPR   & 0.0107 & 0.0043 & 0.0036 & 0.0004 & 0.5832 \\
INLA  & 0.0247 & 0.0132 & 0.0031 & 0.0003 & 2.0533 \\
\hline
\end{tabular}
\caption{Average performance over $50$ simulated replicates under $10\%$ location-wise missingness: mean squared error of the estimated category probabilities ($\mathrm{MSE}_{\pi}$) at the held-out (missing) and observed locations, and mean elapsed computation time in seconds. The INLA timing includes drawing $1{,}000$ posterior samples.}
\label{tab:sim_main}
\end{table}

At the observed locations, the three methods perform similarly, with average probability MSEs between $0.0031$ and $0.0050$. Differences emerge at the held-out locations, where prediction must rely on the covariate and the spatial basis alone: the model implemented with EPR attains an average held-out MSE of $0.0107$, the smallest of the three methods and less than half that of the model implemented with Gibbs sampler ($0.0266$) with the smallest variability across replicates. In terms of computation, EPR is the fastest of the three methods ($0.5832$ seconds), followed by INLA ($2.0533$ seconds, including posterior sampling) and the Gibbs sampler ($5.1551$ seconds).

Table~\ref{tab:sim_pi_uq} summarizes uncertainty quantification for the category probabilities, reporting the empirical coverage of $95\%$ credible intervals and the average interval width, again separated by held-out and observed locations. The two models implemented with Gibbs and EPR respectively both attain coverage at or above the nominal level, with intervals of comparable width. The model fitted with INLA, in contrast, produces intervals that are roughly half as wide (e.g., $0.3296$ versus $0.5702$--$0.6328$ at held-out locations) but undercovers, with empirical coverage of $0.758$ at held-out locations and $0.886$ at observed locations. Coverage and interval width must therefore be read together: the model that INLA approximates understates posterior uncertainty for $\boldsymbol{\pi}$ in this setting, producing intervals that are too short to attain nominal coverage.

\begin{table}[!htbp]
\centering
\vspace{0.2em} \small
\begin{tabular}{lcccc}
\hline
 & \multicolumn{2}{c}{Held-out locations} & \multicolumn{2}{c}{Observed locations} \\
\cmidrule(lr){2-3}\cmidrule(lr){4-5}
Strategy & Coverage & CI width & Coverage & CI width \\
\hline
Gibbs & 0.967 & 0.6328 & 0.9962 & 0.5637 \\
EPR   & 0.993 & 0.5702 & 0.9997 & 0.5193 \\
INLA  & 0.758 & 0.3296 & 0.8860 & 0.1708 \\
\hline
\end{tabular}
\caption{Empirical coverage of pointwise $95\%$ credible intervals for the category
probabilities and average interval widths, at held-out and observed locations, averaged over
$50$ simulated replicates.}
\label{tab:sim_pi_uq}
\end{table}

Table~\ref{tab:sim_beta2} summarizes estimation and uncertainty quantification for the slope coefficient $\beta_2=-2$. The model fitted with EPR provides the most accurate point estimate, with mean estimate $-2.1733$ and coefficient MSE of $0.0420$, compared to $0.1934$ for the model fitted with Gibbs and $0.0834$ for the model fitted with INLA. For uncertainty quantification, the model fitted with Gibbs attains empirical coverage of $0.96$, close to the nominal $0.95$ level, while the model implemented with EPR attains coverage of $1.00$ with a shorter average interval than Gibbs. Inference arising from INLA attains an empirical coverage of only $0.76$. These results indicate that the independent posterior samples generated by EPR provide accurate coefficient inference and uncertainty quantification while avoiding burn-in and Markov chain convergence assessment.
 
\begin{table}[!htbp]
\centering
\vspace{0.2em} \small
\begin{tabular}{lcccc}
\hline
Strategy & $\hat{\beta}_2$ & $\mathrm{MSE}_{\beta_2}$ & Coverage & CI width \\
\hline
Gibbs & -2.3450 & 0.1934 & 0.96 & 1.4400 \\
EPR   & -2.1733 & 0.0420 & 1.00 & 1.0395 \\
INLA  & -1.7898 & 0.0834 & 0.76 & 0.7233 \\
\hline
\end{tabular}
\caption{Point estimation, mean squared error, empirical coverage, and average credible
interval width for the slope coefficient $\beta_2=-2$, averaged over $50$ simulated replicates
under $10\%$ location-wise missingness.}
\label{tab:sim_beta2}
\end{table}

\subsection{Categorical Case}
\label{sec:sim_cat_results}
Next, we consider the categorical case with $m_i=1$, where each location generates a single observation from one of the $K=5$ categories. This setting most closely matches the real-data application. The simulation design is otherwise identical to that in Section~\ref{sec:sim_design}, with $10\%$ location-wise held-out observations in each
replicate. Tables~\ref{tab:sim_cat_main}--\ref{tab:sim_cat_beta2} summarize the results
over $50$ simulated replicates.

\begin{table}[!htbp]
\centering
\vspace{0.4em} \small
\begin{tabular}{lccccc}
\hline
 & \multicolumn{2}{c}{Held-out locations} & \multicolumn{2}{c}{Observed locations} & \\
\cmidrule(lr){2-3}\cmidrule(lr){4-5}
Strategy  & $\mathrm{MSE}_{\pi}$ mean & $\mathrm{MSE}_{\pi}$ sd
& $\mathrm{MSE}_{\pi}$ mean & $\mathrm{MSE}_{\pi}$ sd & CPU (s) \\
\hline
Gibbs & 0.0178 & 0.0082 & 0.0557 & 0.0051 & 3.5704 \\
EPR   & 0.0178 & 0.0068 & 0.0271 & 0.0037 & 0.4466 \\
INLA  & 0.0387 & 0.0137 & 0.0422 & 0.0109 & 2.0638 \\
\hline
\end{tabular}
\caption{Average performance over $50$ simulated replicates for the categorical case ($m_i=1$) under $10\%$ location-wise missingness: mean squared error of the estimated category probabilities ($\mathrm{MSE}_{\pi}$) at the held-out (missing) and observed
locations, and mean elapsed computation time in seconds. The INLA timing includes drawing $1{,}000$ posterior samples.}
\label{tab:sim_cat_main}
\end{table}

\begin{table}[!htbp]
\centering
\vspace{0.2em} \small
\begin{tabular}{lcccc}
\hline
 & \multicolumn{2}{c}{Held-out locations} & \multicolumn{2}{c}{Observed locations} \\
\cmidrule(lr){2-3}\cmidrule(lr){4-5}
Strategy  & Coverage & CI width & Coverage & CI width \\
\hline
Gibbs & 0.984 & 0.8823 & 0.9744 & 0.8318 \\
EPR   & 0.994 & 0.6023 & 0.9852 & 0.6387 \\
INLA  & 0.986 & 0.9684 & 0.9838 & 0.9363 \\
\hline
\end{tabular}
\caption{Empirical coverage of pointwise $95\%$ credible intervals for the category
probabilities and average interval widths in the categorical case ($m_i=1$), at held-out and
observed locations, averaged over $50$ simulated replicates.}
\label{tab:sim_cat_pi_uq}
\end{table}

\begin{table}[!htbp]
\centering
\vspace{0.2em} \small
\begin{tabular}{lcccc}
\hline
Strategy  & $\hat{\beta}_2$ & $\mathrm{MSE}_{\beta_2}$ & Coverage & CI width \\
\hline
Gibbs & -9.9585 & 64.5452 & 0.00 & 9.9012 \\
EPR   & -1.5372 & 0.2453  & 1.00 & 1.9526 \\
INLA  & -5.8750 & 53.5082 & 0.98 & 20.5908 \\
\hline
\end{tabular}
\caption{Point estimation, mean squared error, empirical coverage, and average credible
interval width for the slope coefficient $\beta_2=-2$ in the categorical case ($m_i=1$),
averaged over $50$ simulated replicates.}
\label{tab:sim_cat_beta2}
\end{table}

Table~\ref{tab:sim_cat_main} shows that the models implemented with Gibbs and EPR have the same average probability MSE of $0.0178$, although EPR has slightly smaller variability across replicates. The model fitted with INLA has a larger held-out MSE of $0.0387$. At the observed locations, EPR produces the smallest prediction error, while Gibbs sampling led to noticeably larger MSE. EPR is also the fastest of the three strategies in this experiment.

The pointwise uncertainty quantification for $\boldsymbol{\pi}$ is reported in Table~\ref{tab:sim_cat_pi_uq}. All three strategies achieve coverage at or above, or close to the nominal $95\%$ level at both held-out and observed locations. However, the model fitted with EPR achieves this coverage with the narrowest credible intervals, with average widths of $0.6023$ at held-out locations and $0.6387$ at observed locations.

Table~\ref{tab:sim_cat_beta2} shows a different pattern for coefficient inference. The models using Gibbs sampler and INLA have substantial bias in estimating $\beta_2=-2$, with posterior means of $-9.9585$ and $-5.8750$, respectively. The model fitted with Gibbs also has poor empirical coverage ($0.00$), while the model implemented with INLA attains high coverage only through very wide intervals.

To assess whether the poor coefficient recovery for Gibbs was attributable to insufficient MCMC convergence, we conducted additional convergence diagnostics using three longer chains for a representative categorical replicate; see Appendix~\ref{app:gibbs-convergence}. The trace plots exhibited stable behavior, and the Gelman--Rubin diagnostic,  $\hat{R}$, for $\beta_2$ was $1.00016$, suggesting that the poor recovery is not explained simply by convergence issues. This difference between the results using a model implemented with Gibbs sampler and another discrepancy model fitted with EPR may be due to the differences in their discrepancy model assumptions.

While the model implemented with EPR also shows some attenuation toward zero ($-1.53$ versus the true $-2$), its bias is far smaller than the models fitted with Gibbs and INLA, and it maintains coverage at or above the nominal level with substantially narrower intervals. Thus, in the categorical case, the model fitted with EPR provides the most stable balance between prediction accuracy, uncertainty quantification, and coefficient inference.

\FloatBarrier
\section{Application}
\label{sec:application}
We analyze land-water transitions in the Great Lakes region using land-cover classifications from the MODIS/Terra+Aqua Land Cover Type Yearly L3 Global 0.05-degree CMG product (MCD12C1, Collection 6.1) for 2010 and 2020 \citep{friedl2022}. The 2010 and 2020 land-cover classifications are used to define four transition categories:
water-to-water (WW), water-to-land (WL), land-to-water (LW), and land-to-land (LL).

\begin{table}[t]
\centering
\begin{tabular}{lrrrr}
\toprule
Category & WW & WL & LW & LL \\
\midrule
Count      & $359{,}699$ & $1{,}473$ & $1{,}008$ & $547{,}820$ \\
Proportion & $0.395$ & $0.0016$ & $0.0011$ & $0.602$ \\
\bottomrule
\end{tabular}
\caption{Distribution of land-cover transition categories in the analysis region.}
\label{tab:transition-counts}
\end{table}

The dataset contains $910{,}000$ spatial locations. $359{,}699$ are classified as WW, $1{,}473$ as WL, $1{,}008$ as LW, and $547{,}820$ as LL. Thus, transitions between water and land are extremely rare, accounting for less than $0.3\%$ of the observations. We call WW and LL ``persistence categories'' as their land classification does not change between 2010 and 2020; similarly, we call WL and LW  ``transition categories".

We fit the multinomial EPR model described in Section ~\ref{sec:process_model} using three stick-breaking layers. The covariate design allows the regression coefficients $\boldsymbol{\beta}_h$ to vary across layers, whereas the spatial basis coefficient vector $\boldsymbol{\eta}$ is shared across the three layers. The spatial effect is represented using $125$ Gaussian radial basis functions, consisting of $25$ regularly spaced knots and $100$ additional knots concentrated near observed transition locations. Posterior inference is based on $B=500$ independent EPR replicates, with the resulting stick-breaking probabilities transformed back to the four-category multinomial probability scale.

For covariates, we use land surface temperature (LST) and precipitation for both 2010 and 2020, motivated by their relationship with Great Lakes water-level variability discussed in Section~\ref{sec:intro}. LST is obtained from the MODIS/Terra Land Surface Temperature/Emissivity Monthly L3 Global 0.05-degree Climate Modeling Grid product (MOD11C3, Collection 6.1) \citep{WanHookHulley2021}, and precipitation is obtained from the ERA5-Land monthly averaged data provided by the Copernicus Climate Change Service Climate Data Store \citep{MunozSabater2019,MunozSabater2021}. All continuous covariates are standardized prior to model fitting, and missingness indicators are included for LST.

\subsection{Results}
\label{sec:app-results}
To evaluate predictive performance, we use a stratified holdout sample of $600$ locations consisting of $200$ WW, $100$ WL, $100$ LW, and $200$ LL observations. The model is fit to the remaining locations, and posterior predictive probabilities are computed for the held-out observations.

Table~\ref{tab:app-auc-shared} reports the resulting predictive AUC values. The persistence categories are predicted well, with AUC $0.9536$ for WW and $0.8059$ for LL. The marginal AUC values for the two rare transition categories are $0.6619$ for WL and $0.6864$ for LW, indicating moderate discrimination between each transition category and the remaining categories.
\begin{table}[!htbp]
\centering
\begin{tabular}{llc}
\toprule
Evaluation & Transition & AUC \\
\midrule
Marginal & Water \(\to\) Water (WW) & $0.9536$ \\
Marginal & Water \(\to\) Land  (WL) & $0.6619$ \\
Marginal & Land  \(\to\) Water (LW) & $0.6864$ \\
Marginal & Land  \(\to\) Land  (LL) & $0.8059$ \\
\midrule
Conditional & W \(\to\) L, given start \(=\) water & $0.9302$ \\
Conditional & L \(\to\) W, given start \(=\) land  & $0.5963$ \\
\bottomrule
\end{tabular}
\caption{Predictive AUC values.}
\label{tab:app-auc-shared}
\end{table}
\FloatBarrier

Because the marginal AUC compares each transition category with all remaining categories, we additionally evaluate AUC conditional on the initial land-water state. For the conditional AUC, we compare locations that share the same initial land-water state. This measures how well the model distinguishes a transition from persistence among locations that began in the same state, rather than distinguishing one category from all other categories. Specifically, among locations that were classified as water in 2010, the model distinguishes WL from WW well, with a conditional AUC of $0.9302$. In contrast, among locations classified as land in 2010, the conditional AUC for distinguishing LW from LL is $0.5963$. Thus, the model is much better at distinguishing water-to-land transition from persistent water than at distinguishing land-to-water transition from persistent land.

For inference on covariate effects, we refit the model to the full dataset of $910{,}000$ locations. Table~\ref{tab:app-beta} reports posterior means and $95\%$ credible intervals from the $B=500$ posterior replicates. Since the continuous covariates are standardized, the coefficient for LST and precipitation represents the change in the conditional log odds associated with a one standard deviation increase in the corresponding covariate. 

\begin{table}[!htbp]
\centering
\small
\begin{tabular}{llrrr}
\toprule
Layer & Covariate & Mean & 2.5\% & 97.5\% \\
\midrule
\multirow{7}{*}{L1}
& Intercept & $-2.7456$ & $-3.6177$ & $-1.9959$ \\
& LST 2010 & $-0.3042$ & $-0.3835$ & $-0.2328$ \\
& LST missing 2010 & $1.9292$ & $1.3079$ & $2.5879$ \\
& LST 2020 & $0.1499$ & $0.0735$ & $0.2308$ \\
& LST missing 2020 & $1.5724$ & $0.9265$ & $2.1894$ \\
& Precipitation 2010 & $0.1815$ & $0.1540$ & $0.2078$ \\
& Precipitation 2020 & $-0.0876$ & $-0.1138$ & $-0.0610$ \\
\midrule

\multirow{7}{*}{L2}
& Intercept & $-3.1566$ & $-4.0124$ & $-2.4039$ \\
& LST 2010 & $0.0090$ & $-0.0832$ & $0.0865$ \\
& LST missing 2010 & $0.0194$ & $-4.5131$ & $5.3994$ \\
& LST 2020 & $0.0265$ & $-0.0582$ & $0.1168$ \\
& LST missing 2020 & $0.0301$ & $-3.5497$ & $4.3027$ \\
& Precipitation 2010 & $0.0631$ & $0.0298$ & $0.0971$ \\
& Precipitation 2020 & $-0.0510$ & $-0.0834$ & $-0.0191$ \\
\midrule

\multirow{7}{*}{L3}
& Intercept & $-3.1596$ & $-4.0228$ & $-2.4165$ \\
& LST 2010 & $-0.0008$ & $-0.0919$ & $0.0788$ \\
& LST missing 2010 & $-0.2587$ & $-5.1451$ & $3.2422$ \\
& LST 2020 & $0.0376$ & $-0.0494$ & $0.1261$ \\
& LST missing 2020 & $-0.1218$ & $-4.1754$ & $3.2539$ \\
& Precipitation 2010 & $0.0654$ & $0.0310$ & $0.0951$ \\
& Precipitation 2020 & $-0.0502$ & $-0.0833$ & $-0.0157$ \\
\bottomrule
\end{tabular}
\caption{Posterior means and $95\%$ credible intervals for the layer-specific regression coefficients with $B=500$ posterior replicates. L1 corresponds to WW versus the remaining categories, L2 corresponds to WL conditional on not WW, and L3 corresponds to LW conditional on neither WW nor WL.}
\label{tab:app-beta}
\end{table}

The estimated covariate effects differ across the three sequential stick-breaking comparisons. In the first comparison, which contrasts WW with the remaining categories, higher 2010 land surface temperature 
(LST) is associated with lower conditional odds of WW, whereas higher 2020 LST is associated with greater conditional odds. Precipitation also shows clear associations in this comparison, with a positive coefficient for 2010 and a negative coefficient for 2020. The corresponding $95\%$ credible intervals exclude zero.

In the second and third layers, the LST coefficients are close to zero and their credible intervals contain zero, providing little evidence of additional LST associations in these conditional models. In contrast, the precipitation effects are consistent across both layers: 2010 precipitation has a positive association, whereas 2020 precipitation has a negative association, with all corresponding credible intervals excluding zero. The LST missingness indicators are more precisely estimated in the first layer but have substantially wider posterior intervals in the second and third layers. Overall, the estimated covariate associations vary across the three stick-breaking layers, particularly for LST.

\begin{figure}[!htbp]
\centering
\includegraphics[width=0.85\textwidth]{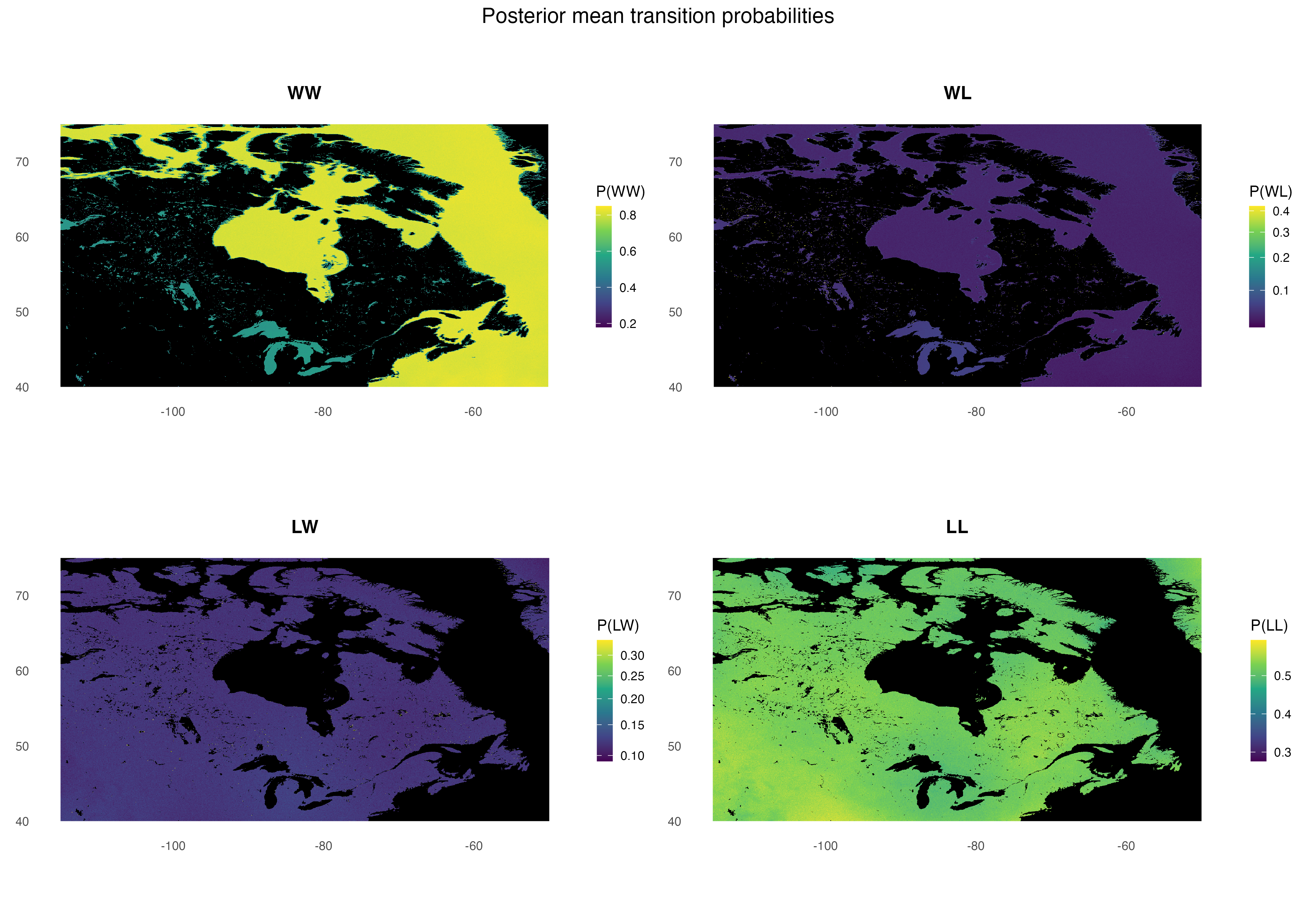}
\caption{Posterior mean probabilities for the four land-water transition categories. Black regions indicate locations where the corresponding transition is not applicable given the initial land-cover state. Each panel uses a category-specific probability scale.}
\label{fig:app-pred-mean}
\end{figure}

Figure~\ref{fig:app-pred-mean} shows the posterior mean probabilities for the four land-water transition categories over the study region. The persistence categories dominate the predicted spatial pattern. Locations initially classified as water generally have high probabilities of remaining water, while locations initially classified as land have relatively high probabilities of remaining land. In contrast, the predicted probabilities of WL and LW are substantially smaller and exhibit spatial variation across the study region. These patterns are consistent with the rarity of the observed transition categories. Black regions indicate locations for which the corresponding transition is not applicable given the initial land-cover state. Because each panel uses a different probability scale, the colors should be interpreted within each category rather than compared directly across panels.

\begin{figure}[!htbp]
\centering
\includegraphics[width=0.85\textwidth]{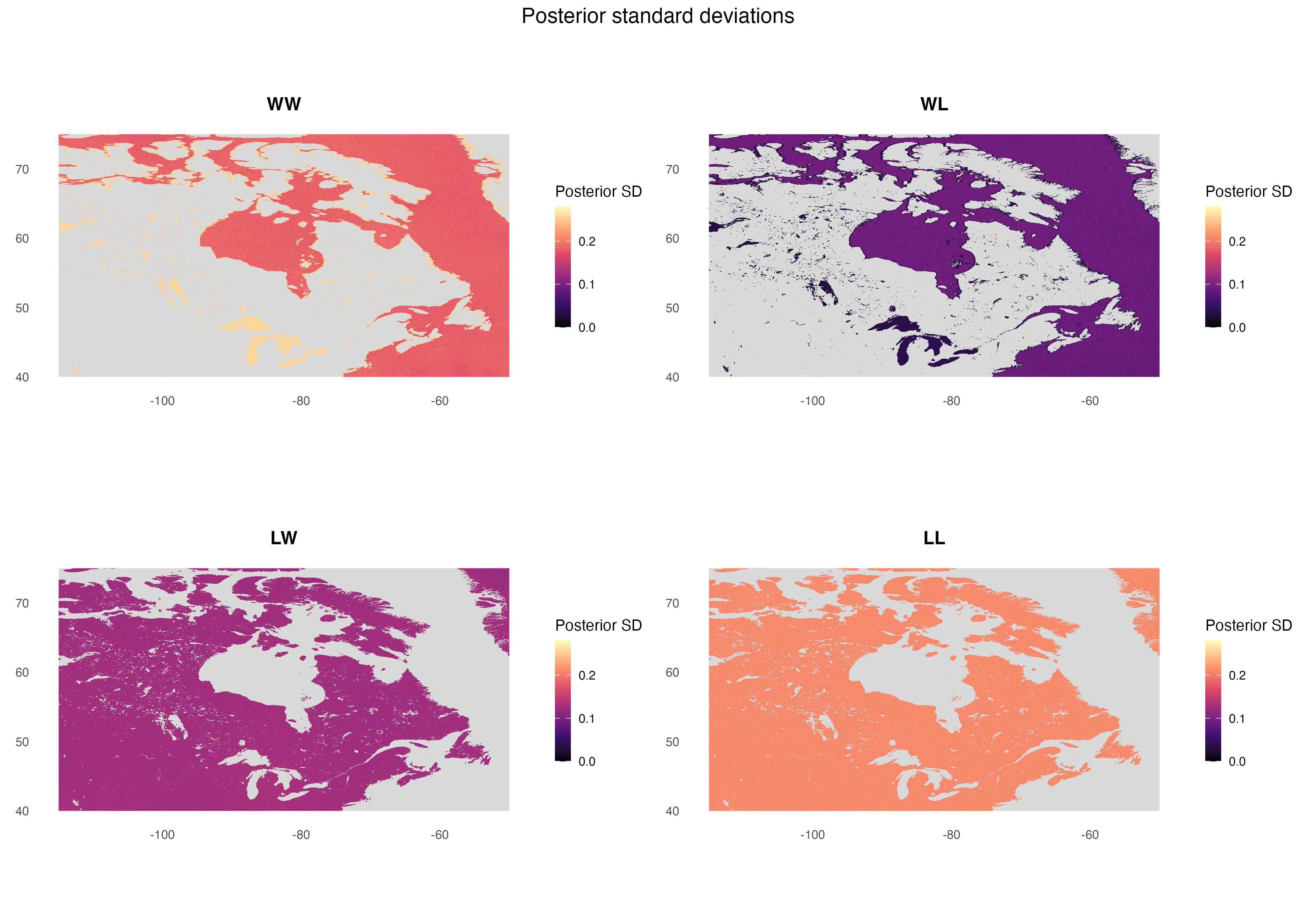}
\caption{Posterior standard deviations of the predicted probabilities
for the four land-water transition categories. Gray regions indicate
locations where the corresponding transition is not applicable given
the initial land-cover state. A common color scale is used across all four panels.}
\label{fig:app-pred-sd}
\end{figure}

Figure~\ref{fig:app-pred-sd} displays the corresponding posterior standard deviations. Posterior uncertainty varies across both space and transition category. The WW and LL probabilities exhibit larger posterior standard deviations over many of their applicable locations, whereas the transition categories generally show smaller posterior standard deviations. A common color scale is used across all four panels to facilitate comparison of posterior uncertainty across categories.

The full data analysis was completed in approximately $47$ minutes. After the stick-breaking expansion, $2{,}009{,}129$ conditional Bernoulli observations contributed to the likelihood.

\section{Discussion}
\label{sec:discussion}
In this paper, we extend EPR to multinomial spatial data using a stick-breaking representation and apply the resulting model to land-water transitions in the Great Lakes region. Our first contribution is methodological. We use the stick-breaking representation to express the multinomial likelihood as a sequence of conditional binomial likelihoods, which allows us to adapt EPR to the multinomial setting and generate independent samples directly from the posterior distribution without MCMC. The simulation results show that our model fitted with EPR achieves competitive predictive accuracy and reliable uncertainty quantification while requiring substantially less computation than other computationally motivated models implemented with Gibbs sampling and INLA. These advantages were also observed in the categorical setting. Our second contribution is theoretical. We characterized the posterior cross-covariance between the latent signal and the additive noise under the multinomial stick-breaking model, extending previous signal-to-noise dependence results to the multinomial setting. Our third contribution is empirical. We applied the multinomial EPR model to $910{,}000$ locations in the Great Lakes region to study land-water transitions from 2010 to 2020. Conditional on locations classified as water in 2010, the model distinguished water-to-land transitions from persistent water well, with an AUC of $0.9302$. We also found spatial variation in the transition probabilities and different associations of land surface temperature and precipitation across the three conditional binomial models induced by the stick-breaking representation.

One limitation of the current analysis is that we consider land-water transitions between only two time points, 2010 and 2020. We selected a ten-year interval because larger changes in land-water classification were observed over this period, but the estimated transition patterns may differ over other time points and on temporal scales. A direction for future work is therefore to extend the model to a spatio-temporal setting with multiple time points and possibly on finer time scales. This would allow land-water transitions to be studied over time and could also allow future transitions to be considered through posterior predictive inference.

\section{Proof of Proposition~\ref{prop:gcm_epr}}
\label{app:proof_gcm_epr}

Using the notation introduced in Section~\ref{sec:post_gcm}, let $M=2N+p+r$, $\mathbf{V}_M^{-1}=(\mathbf{H},\mathbf{Q})$, and
$\mathbf{D}(\theta)^{-1}
=\operatorname{blkdiag}
(\mathbf{I}_N, \mathbf{D}_{\beta}(\theta)^{-1}, \mathbf{D}_{\eta}(\theta)^{-1}, \sigma_{\xi}^{-1}\mathbf{I}_N)$.
Define $\boldsymbol{\alpha}_{M}=\bigl(\mathbf{y}'+\alpha_{\xi}\mathbf{1}_{N}',\, \mathbf{0}_{1,N+p+r}\bigr)'$ and $\boldsymbol{\kappa}_{M}=\bigl(\mathbf{n}'+2\alpha_{\xi}\mathbf{1}_{N}',\,\tfrac12\mathbf{1}_{1,N+p+r}\bigr)'$. For $\mathbf{h}
=(h_1,\ldots,h_N,h_1^*,\ldots,h_{N+p+r}^*)'$, let $\boldsymbol{\psi}_M(\mathbf{h})=\bigl(\psi_{3}(h_{1}),\ldots,\psi_{3}(h_{N}),\psi_{1}(h_{1}^{*}),\ldots,\psi_{1}(h_{N+p+r}^{*})\bigr)'$. Let $\pi(\theta)$ denote the prior density of $\theta$. The data model can be written as
\begin{equation}\label{eq:datamodel}
  f(\mathbf{y}\mid\bm{\xi},\bm{\beta},\bm{\eta},\bm{\theta},\mathbf{q})\;= \; N
  \exp\!\big[\mathbf{y}'\{(\mathbf{I}_N,\mathbf{X},\mathbf{G})\bm{\zeta}-\bm{\delta}_\nu\}
  -\mathbf{n}'\bm{\psi}_3\{(\mathbf{I}_N,\mathbf{X},\mathbf{G})\bm{\zeta}-\bm{\delta}_\nu\}\big],
\end{equation}
where $\bm{\psi}_3(\mathbf{h})=(\log\{1+e^{h_1}\},\dots,\log\{1+e^{h_N}\})'$, $N = \prod_{k,i}\binom{n_{ki}}{Y_{ki}}$ in the binomial setting, $\bm{\zeta} = (\boldsymbol{\xi}', \bm{\beta}', \bm{\eta}')'$, and $\bm{\delta_{\nu}}$ is a function of $\bm{\theta}$ and $\bm{q}$. We want to show that $f(\bm{\zeta},\mathbf{q}|\mathbf{y})\propto \int_\Omega\bm{\pi(\bm{\theta})}\,f(\bm{\zeta},\mathbf{q},\mathbf{y}|\bm{\theta})\,d\bm{\theta}$ is the stated GCM, where the natural parameter is $\bm{\nu}=(\mathbf{I}_N,\mathbf{X},\mathbf{G})\bm{\zeta}-\bm{\delta}_\nu = \mathbf{X}\bm{\beta} + \mathbf{G}\bm{\eta} + \bm{\xi} - \bm{\delta_{\nu}}$. The density $f(\bm{\zeta},\mathbf{q},\bm{\theta}\mid\mathbf{y})$, 
\begin{equation}\label{eq:bayes}
  f(\bm{\zeta},\mathbf{q},\bm{\theta}\mid\mathbf{y})\;\propto\;
  f(\mathbf{y}\mid\bm{\xi},\bm{\beta},\bm{\eta},\bm{\theta},\mathbf{q})\,
  f(\bm{\xi}\mid\bm{\beta},\bm{\eta},\bm{\theta},\mathbf{q})\,
  f(\bm{\beta}\mid\bm{\theta},\mathbf{q})\,
  f(\bm{\eta}\mid\bm{\theta},\mathbf{q})\,
  f(\mathbf{q})\,\bm{\pi(\bm{\theta})},
\end{equation}
where $f(\mathbf{q})=1$ and $\bm{\delta} = (\bm{\delta_{\nu}}', \bm{\delta_{\beta}}', \bm{\delta_{\eta}}', \bm{\delta_{\xi}}')' = -\boldsymbol{D(\theta)^{-1}\bm{Q}\bm{q}}$. Then,

\begin{equation}
\label{eq:finescale}
\begin{aligned}
  f(\bm{\xi}\mid\bm{\beta},\bm{\eta},\bm{\theta},\mathbf{q})
  &\propto
  \exp\Bigg[
  \bm{\alpha}_\xi'
  \left\{
  \begin{pmatrix}
  \mathbf{I}_{N} & \mathbf{X} & \mathbf{G}\\[1pt]
  \sigma_\xi^{-1}\mathbf{I}_{N} & \mathbf{0}_{N,p} & \mathbf{0}_{N,r}
  \end{pmatrix}
  \bm{\zeta}
  -
  \begin{pmatrix}
  \bm{\delta}_{\nu}\\[1pt]
  \bm{\delta}_{\xi}
  \end{pmatrix}
  \right\} \\
  &\qquad
  -
  \bm{\kappa}_\xi'
  \bm{\psi}_{D,\xi}
  \left\{
  \begin{pmatrix}
  \mathbf{I}_{N} & \mathbf{X} & \mathbf{G}\\[1pt]
  \sigma_\xi^{-1}\mathbf{I}_{N} & \mathbf{0}_{N,p} & \mathbf{0}_{N,r}
  \end{pmatrix}
  \bm{\zeta}
  -
  \begin{pmatrix}
  \bm{\delta}_{\nu}\\[1pt]
  \bm{\delta}_{\xi}
  \end{pmatrix}
  \right\}
  \Bigg].
\end{aligned}
\end{equation}
with $\sigma_{\xi}^2$ and $\alpha_\xi$ known, $\bm{\alpha}_{\xi} = (\alpha_{\xi}\mathbf{1}_{1,N}, \mathbf{0}_{1,N})'$, $\bm{\kappa}_{\xi} = (2\alpha_{\xi}\mathbf{1}_{1,N}, \frac{1}{2}\mathbf{1}_{1,N})'$. Here, $\bm{\psi}_{3}$ denotes the componentwise unit log-partition function associated with the binomial likelihood. Recall that $\bm{\delta_{\nu}}, \bm{\delta_{\xi}}$ are functions of $\bm{\theta}$ and $\bm{q}$. Let $\bm{\psi}_{D, \xi}(\bm{h}_{D, \xi}) = (\bm{\psi}_{3}(\bm{h})', \bm{\psi}_{1}(\bm{h^{*}})')'$ for $2N$-dimensional real-valued vector $\bm{h}_{D, \xi} = (\bm{h}', (\bm{h^{*}})')'$, where $\psi_1(h)=h^2$ and $\psi_3(h)=\log\{1+\exp(h)\}$.

The product, 
\begin{equation}
\label{eq:combined}
\begin{aligned}
& f(\mathbf{y}\mid\bm{\xi},\bm{\beta},\bm{\eta},\bm{\theta},\mathbf{q})
  f(\bm{\xi}\mid\bm{\beta},\bm{\eta},\bm{\theta},\mathbf{q})  \\
&\quad \propto
    \exp \Bigg[\bm{a}_{y}'
    \left\{
    \begin{pmatrix}
    \mathbf{I}_{N} & \mathbf{X} & \mathbf{G}\\[1pt]
    \sigma_\xi^{-1}\mathbf{I}_{N} & \mathbf{0}_{N,p} & \mathbf{0}_{N,r}
    \end{pmatrix}
    \bm{\zeta}
    -
    \begin{pmatrix}
    \bm{\delta}_{\nu}\\[1pt]
    \bm{\delta}_{\xi}
    \end{pmatrix}
    \right\} \\
&\qquad
    -
    \bm{b}_{y}'
    \bm{\psi}_{D,\xi}
    \left\{
    \begin{pmatrix}
    \mathbf{I}_{N} & \mathbf{X} & \mathbf{G}\\[1pt]
    \sigma_\xi^{-1}\mathbf{I}_{N} & \mathbf{0}_{N,p} & \mathbf{0}_{N,r}
    \end{pmatrix}
    \bm{\zeta}
    -
    \begin{pmatrix}
    \bm{\delta}_{\nu}\\[1pt]
    \bm{\delta}_{\xi}
    \end{pmatrix}
    \right\}
\Bigg].
\end{aligned}
\end{equation}
where $\mathbf{a}_{\mathbf{y}}=(\mathbf{y}'+\alpha_\xi\mathbf{1}_{1,N}, \mathbf{0}_{1,N})'$ and $\mathbf{b}_{\mathbf{y}}=(\mathbf{n}'+2\alpha_\xi\mathbf{1}_{1,N}, \tfrac12\mathbf{1}_{1,N})'$.

Next, the remaining priors $\bm{\beta}|\theta,\mathbf{q}\sim N(\mathbf{D}_\beta\bm{\delta}_\beta,\mathbf{D}_\beta\mathbf{D}_\beta')$ and $\bm{\eta}|\theta,\mathbf{q}\sim N(\mathbf{D}_\eta\bm{\delta}_\eta,\mathbf{D}_\eta\mathbf{D}_\eta')$
are themselves Diaconis--Ylvisaker kernels. Then,
$$
f(\bm{\beta}|\bm{\theta}, \bm{q}) \propto \exp[-\frac{1}{2}\mathbf{1}_{1,p}\psi_1(\mathbf{D}_{\beta}(\bm{\theta})^{-1}\bm{\beta} - \bm{\delta_{\beta}})],
$$
with shape $(\mathbf{0}_{1,p},\tfrac12\mathbf{1}_{1,p})$, and $\psi_1(\cdot) = (\cdot)^2$. The same holds for $\bm{\eta}$ with $p$ replaced by $r$. Expressed through $\zeta$, their arguments are 
$(\mathbf{0}_{p,N},\mathbf{D}_\beta(\theta)^{-1},\mathbf{0}_{p,r})\bm{\zeta}-\bm{\delta}_\beta$ and
$(\mathbf{0}_{r,N},\mathbf{0}_{r,p},\mathbf{D}_\eta(\theta)^{-1})\bm{\zeta}-\bm{\delta}_\eta$.

Multiplying by $f(\bm{\beta}|\bm{\theta},\bm{q})f(\bm{\eta}|\bm{\theta},\bm{q})f(\mathbf{q})\bm{\pi}(\bm{\theta})$, and stacking vectors and matrices leads to 
\begin{equation}
\label{eq:posterior_gcm_expanded}
\begin{aligned}
f(\bm{\zeta},\mathbf q,\bm{\theta}| \mathbf y)
&\propto
\frac{\bm{\pi(\theta)}}{\bm{\det\{\mathbf{V}_M}\}}N    
\exp\Bigg[
\bm{\alpha}_M'
\left\{
\begin{pmatrix}
\mathbf I_N & \mathbf X & \mathbf G\\
\mathbf 0_{p,N} & \mathbf D_\beta(\theta)^{-1} & \mathbf 0_{p,r}\\
\mathbf 0_{r,N} & \mathbf 0_{r,p} & \mathbf D_\eta(\theta)^{-1}\\
\sigma_\xi^{-1}\mathbf I_N & \mathbf 0_{N,p} & \mathbf 0_{N,r}
\end{pmatrix}
\bm{\zeta}
-
\begin{pmatrix}
\bm{\delta}_{\nu}\\
\bm{\delta}_{\beta}\\
\bm{\delta}_{\eta}\\
\bm{\delta}_{\xi}
\end{pmatrix}
\right\} \\
&\hspace{2.7cm}
-
\bm{\kappa}_M'
\bm{\psi}_M
\left(
\left\{
\begin{pmatrix}
\mathbf I_N & \mathbf X & \mathbf G\\
\mathbf 0_{p,N} & \mathbf D_\beta(\theta)^{-1} & \mathbf 0_{p,r}\\
\mathbf 0_{r,N} & \mathbf 0_{r,p} & \mathbf D_\eta(\theta)^{-1}\\
\sigma_\xi^{-1}\mathbf I_N & \mathbf 0_{N,p} & \mathbf 0_{N,r}
\end{pmatrix}
\bm{\zeta}
-
\begin{pmatrix}
\bm{\delta}_{\nu}\\
\bm{\delta}_{\beta}\\
\bm{\delta}_{\eta}\\
\bm{\delta}_{\xi}
\end{pmatrix}
\right\}
\right)
\Bigg].
\end{aligned}
\end{equation} 

Since $\mathbf{D}(\theta)^{-1}\mathbf{H}\bm{\zeta} - \boldsymbol{\delta} =
\mathbf{D}(\theta)^{-1}\mathbf{H}\bm{\zeta} + \mathbf{D}(\theta)^{-1}\mathbf{Q}\mathbf{q} = 
\mathbf{D}(\theta)^{-1} (\mathbf{H}, \mathbf{Q})
\begin{pmatrix}\bm{\zeta}\\\mathbf{q}\end{pmatrix}$, we can integrate with respect to $\bm{\theta}$. Therefore,  
\begin{equation}
\label{eq:gcm_posterior_kernel}
\begin{aligned}
f(\bm{\zeta},\mathbf{q}\mid\mathbf{y})
&\propto
\frac{\mathcal{C}_{M}(\bm{\alpha}_{M},\bm{\kappa}_{M})}{\bm{\det\{\mathbf{V}_M}\}}
\int_{\Omega}
\frac{
\bm{\pi(\theta)}\,
}{
\bm{\det\{\mathbf{D}(\theta)\}}
}
\exp\Bigg[
\bm{\alpha}_{M}'
\mathbf{D}(\bm{\theta})^{-1}
(\mathbf{H}, \mathbf{Q})
\begin{pmatrix}
\bm{\zeta}\\
\mathbf{q}
\end{pmatrix}  \\
&\qquad\qquad
-
\bm{\kappa}_{M}'
\bm{\psi}_{M}
\left\{
\mathbf{D}(\theta)^{-1}
(\mathbf{H}, \mathbf{Q})
\begin{pmatrix}
\bm{\zeta}\\
\mathbf{q}
\end{pmatrix}
\right\}
\Bigg]
\,d\bm{\theta} \\
&\propto 
\mathrm{GCM}\!\big(\bm{\alpha}_M,\bm{\kappa}_M,\mathbf{0}_M,\mathbf{V}_M,\bm{\pi},\mathbf{D};\bm{\psi}_M\big),
\end{aligned}
\end{equation}
where $\mathcal{C}_{M}(\bm{\alpha}_{M},\bm{\kappa}_{M}) = (2\pi)^{-(N+p+r)/2} \Pi_{i=1}^{N}\frac{\Gamma(n_{ki} + 2\alpha_{\xi})}{\Gamma(Y_{ki} + \alpha_{\xi})\Gamma(n_{ki} - Y_{ki} + \alpha_{\xi})}$, and $M=2N+p+r$,
which proves the proposition.

\section{Proof of Proposition~\ref{prop:signal_noise}}
\label{app:proof_signal_noise}
Let $\mathbf{P}_{H}
=\mathbf{H}(\mathbf{H}'\mathbf{H})^{-1}\mathbf{H}'$.
From the GCM/EPR representation in
Proposition~\ref{prop:gcm_epr},
\[
\boldsymbol{\zeta}
=
(\mathbf{H}'\mathbf{H})^{-1}
\mathbf{H}'\mathbf{D}(\theta)\boldsymbol{w}_{M},
\qquad
\mathbf{q}
=
\mathbf{Q}'\mathbf{D}(\theta)\boldsymbol{w}_{M}.
\]
Multiplying the first equation by \(\mathbf{H}\) gives
\[
    \mathbf{H}\boldsymbol{\zeta}
    =
    \mathbf{H}
    (\mathbf{H}'\mathbf{H})^{-1}
    \mathbf{H}'
    \mathbf{D}(\theta)\boldsymbol{w}_{M}
    =
    \mathbf{P}_{H}\mathbf{D}(\theta)\boldsymbol{w}_{M}.
\]
Since the first block of $\mathbf{H}\boldsymbol{\zeta}$ is
$\boldsymbol{\nu}^{*}$, and $\mathbf{J}_{\nu} = (\mathbf{I}_N,\bm{0}_{N,M-N})$ is the matrix that selects the first $N$ entries of an $M$-dimensional vector, we have $\boldsymbol{\nu}^{*}=
\mathbf{J}_{\nu} \mathbf{P}_{H} \mathbf{D}(\theta) \boldsymbol{w}_{M}$. Hence, by the definition of the discrepancy term $\boldsymbol{\delta} = - \mathbf{D}(\theta)^{-1} \mathbf{Q}\mathbf{q}$. Since the first block of $\mathbf{D}(\theta)^{-1}$ is $\mathbf{I}_{N}$,
$-\boldsymbol{\delta}_{\nu} = \mathbf{J}_{\nu}\mathbf{Q}\mathbf{q}$.
Using $\mathbf{q} = \mathbf{Q}'\mathbf{D}(\theta)\boldsymbol{w}_{M}$
and $\mathbf{Q}\mathbf{Q}' = \mathbf{I}_{M}-\mathbf{P}_{H}$, we obtain $-\boldsymbol{\delta}_{\nu}
= \mathbf{J}_{\nu} (\mathbf{I}_{M}-\mathbf{P}_{H}) \mathbf{D}(\theta)\boldsymbol{w}_{M}$.

Therefore,
\[
\begin{aligned}
    &\operatorname{Cov}
    \left(
    \boldsymbol{\nu}^{*},
    -\boldsymbol{\delta}_{\nu}
    \mid
    \mathbf{y},\theta
    \right) \\
    &=
    \operatorname{Cov}
    \left(
    \mathbf{J}_{\nu}
    \mathbf{P}_{H}
    \mathbf{D}(\theta)
    \boldsymbol{w}_{M},
    \mathbf{J}_{\nu}
    (\mathbf{I}_{M}-\mathbf{P}_{H})
    \mathbf{D}(\theta)
    \boldsymbol{w}_{M}
    \mid
    \mathbf{y},\theta
    \right) \\
    &=
    \mathbf{J}_{\nu}
    \mathbf{P}_{H}
    \mathbf{D}(\theta)
    \operatorname{Cov}
    \left(
    \boldsymbol{w}_{M}
    \mid
    \boldsymbol{\alpha}_{M},
    \boldsymbol{\kappa}_{M}
    \right)
    \mathbf{D}(\theta)'
    (\mathbf{I}_{M}-\mathbf{P}_{H})
    \mathbf{J}_{\nu}' \\
    &=
    \mathbf{J}_{\nu}
    \mathbf{H}
    (\mathbf{H}'\mathbf{H})^{-1}
    \mathbf{H}'
    \mathbf{D}(\theta)
    \operatorname{Cov}
    \left(
    \boldsymbol{w}_{M}
    \mid
    \boldsymbol{\alpha}_{M},
    \boldsymbol{\kappa}_{M}
    \right)
    \mathbf{D}(\theta)'
    \left\{
    \mathbf{I}_{M}
    -
    \mathbf{H}
    (\mathbf{H}'\mathbf{H})^{-1}
    \mathbf{H}'
    \right\}
    \mathbf{J}_{\nu}'.
\end{aligned}
\]
This gives the conditional cross-covariance, given $\mathbf{y}$ and $\theta$, stated in Proposition~\ref{prop:signal_noise}. Since the right-hand side is not constrained to be zero in general, the proposed model permits posterior dependence between the latent stick-breaking signal and the additive noise
contribution. 

We now consider the covariance marginally over the data $\mathbf{y}$, with $\theta$
held fixed. From the conditional derivation, $\boldsymbol{\nu}^{*} = \mathbf{J}_{\nu}\mathbf{P}_{H}\mathbf{D}(\theta)\boldsymbol{w}_{M}$, and $-\boldsymbol{\delta}_{\nu} = \mathbf{J}_{\nu}(\mathbf{I}_{M}-\mathbf{P}_{H})\mathbf{D}(\theta)\boldsymbol{w}_{M}$, in which the matrices $\mathbf{J}_{\nu}\mathbf{P}_{H}\mathbf{D}(\theta)$ and
$\mathbf{J}_{\nu}(\mathbf{I}_{M}-\mathbf{P}_{H})\mathbf{D}(\theta)$ depend only on $\theta$ and the
fixed $\mathbf{J}_{\nu},\mathbf{P}_{H}$, and not on the data $\mathbf{y}$ or on
$\boldsymbol{w}_{M}$. Given $\theta$ they are therefore constant, so by bilinearity of covariance
\begin{equation}
    \operatorname{Cov}\!\left(\boldsymbol{\nu}^{*},-\boldsymbol{\delta}_{\nu}\mid\theta\right)
    =\mathbf{J}_{\nu}\mathbf{P}_{H}\mathbf{D}(\theta)\,
    \operatorname{Cov}\!\left(\boldsymbol{w}_{M}\mid\theta\right)\,
    \mathbf{D}(\theta)'(\mathbf{I}_{M}-\mathbf{P}_{H})\mathbf{J}_{\nu}'.
\end{equation}
Writing $\boldsymbol{\mu}_{w}(\mathbf{y}):=E[\boldsymbol{w}_{M}\mid\mathbf{y}]$ and
$\boldsymbol{\Sigma}_{w}(\mathbf{y}):=\operatorname{Cov}(\boldsymbol{w}_{M}\mid\mathbf{y})$, and
decomposing over the data by the law of total covariance, 
\begin{equation}
    \operatorname{Cov}\!\left(\boldsymbol{w}_{M}\mid\theta\right)
    =E_{\mathbf{y}}\!\big[\operatorname{Cov}(\boldsymbol{w}_{M}\mid\mathbf{y})\big]
    +\operatorname{Cov}_{\mathbf{y}}\!\big(E[\boldsymbol{w}_{M}\mid\mathbf{y}]\big)
    =E_{\mathbf{y}}\!\big[\boldsymbol{\Sigma}_{w}(\mathbf{y})\big]
    +\operatorname{Cov}_{\mathbf{y}}\!\big(\boldsymbol{\mu}_{w}(\mathbf{y})\big),
\end{equation}
so that
\begin{equation}
    \operatorname{Cov}\!\left(\boldsymbol{\nu}^{*},-\boldsymbol{\delta}_{\nu}\mid\theta\right)
    =\mathbf{J}_{\nu}\mathbf{P}_{H}\mathbf{D}(\theta)
     \Big\{
       E_{\mathbf{y}}\!\big[\boldsymbol{\Sigma}_{w}(\mathbf{y})\big]
       +\operatorname{Cov}_{\mathbf{y}}\!\big(\boldsymbol{\mu}_{w}(\mathbf{y})\big)
     \Big\}
     \mathbf{D}(\theta)'(\mathbf{I}_{M}-\mathbf{P}_{H})\mathbf{J}_{\nu}'.
\end{equation}

The conditional moments $\boldsymbol{\mu}_{w}(\mathbf{y})$ and $\boldsymbol{\Sigma}_{w}(\mathbf{y})$
are closed form the moment generating function of the Diaconis--Ylvisaker family
\citep[Eqs.~(3)--(5)]{BHW2020}. The normalizing constant for the logit-beta case
$\psi(w)=\log\{1+\exp(w)\}$ is derived via the multivariate logit-beta distribution
\citep{WB2024}, and the resulting moments are expressed in terms of the digamma and
trigamma functions as detailed below.

Specializing to the logit-beta case $\psi_3(w)=\log\{1+\exp(w)\}$, the change of variables
$p=e^{w}/(1+e^{w})$, for which $dp=p(1-p)\,dw$, gives
\[
  K(\alpha,\kappa)^{-1}
  =\int_{-\infty}^{\infty}\frac{e^{\alpha w}}{(1+e^{w})^{\kappa}}\,dw
  =\int_{0}^{1}p^{\alpha-1}(1-p)^{\kappa-\alpha-1}\,dp
  =\frac{\Gamma(\alpha)\Gamma(\kappa-\alpha)}{\Gamma(\kappa)},
\]
hence $K(\alpha,\kappa)=\Gamma(\kappa)/\{\Gamma(\alpha)\Gamma(\kappa-\alpha)\}$.

Since $\log K(\alpha,\kappa)=\log\Gamma(\kappa)-\log\Gamma(\alpha)-\log\Gamma(\kappa-\alpha)$ and
$\partial_{\alpha}(\kappa-\alpha)=-1$,
\[
  E[w]=-\frac{d}{d\alpha}\log K(\alpha,\kappa)
  =\psi^{(0)}(\alpha)-\psi^{(0)}(\kappa-\alpha),
\]
\[
  \operatorname{Var}(w)=-\frac{d^{2}}{d\alpha^{2}}\log K(\alpha,\kappa)
  =\psi^{(1)}(\alpha)+\psi^{(1)}(\kappa-\alpha).
\]

Substituting $\alpha=Y_{ki}+\alpha_{\xi}$ and $\kappa-\alpha=n_{ki}-Y_{ki}+\alpha_{\xi}$,
\[
  [\boldsymbol{\mu}_{w}(\mathbf{y})]_{i}
  =\psi^{(0)}\!\left(Y_{ki}+\alpha_{\xi}\right)
  -\psi^{(0)}\!\left(n_{ki}-Y_{ki}+\alpha_{\xi}\right),
  \quad
  [\boldsymbol{\Sigma}_{w}(\mathbf{y})]_{ii}
  =\psi^{(1)}\!\left(Y_{ki}+\alpha_{\xi}\right)
  +\psi^{(1)}\!\left(n_{ki}-Y_{ki}+\alpha_{\xi}\right),
\]
where $\psi^{(0)},\psi^{(1)}$ are the digamma and trigamma functions. The remaining $N+p+r$ components of $\boldsymbol{w}_{M}$, corresponding to $\psi_1(w)=w^{2}$ with $\alpha=0$ and $\kappa=\tfrac{1}{2}$, have mean $0$ and variance $1$. The first term averages the conditional covariance matrices, while the second term captures variation in the conditional mean $\boldsymbol{\mu}_{w}(\mathbf{y})$ across $\mathbf{y}$. Neither term is constrained
to equal zero, so the central matrix need not be proportional to $\mathbf{I}_{M}$. Hence, although $\mathbf{P}_{H}(\mathbf{I}_{M}-\mathbf{P}_{H})=\mathbf{0}$, the marginal cross-covariance need not equal zero.

\section{Gibbs Sampler Convergence Diagnostics}
\label{app:gibbs-convergence}

To assess whether the poor recovery of $\beta_2$ in the categorical simulation setting was attributable to insufficient MCMC convergence, we conducted an additional convergence analysis for a representative simulation replicate. Three Gibbs chains were run for 20,000 iterations each, with the first 10,000 iterations discarded as burn-in.

\begin{figure}[ht]
    \centering
    \includegraphics[width=0.8\textwidth]{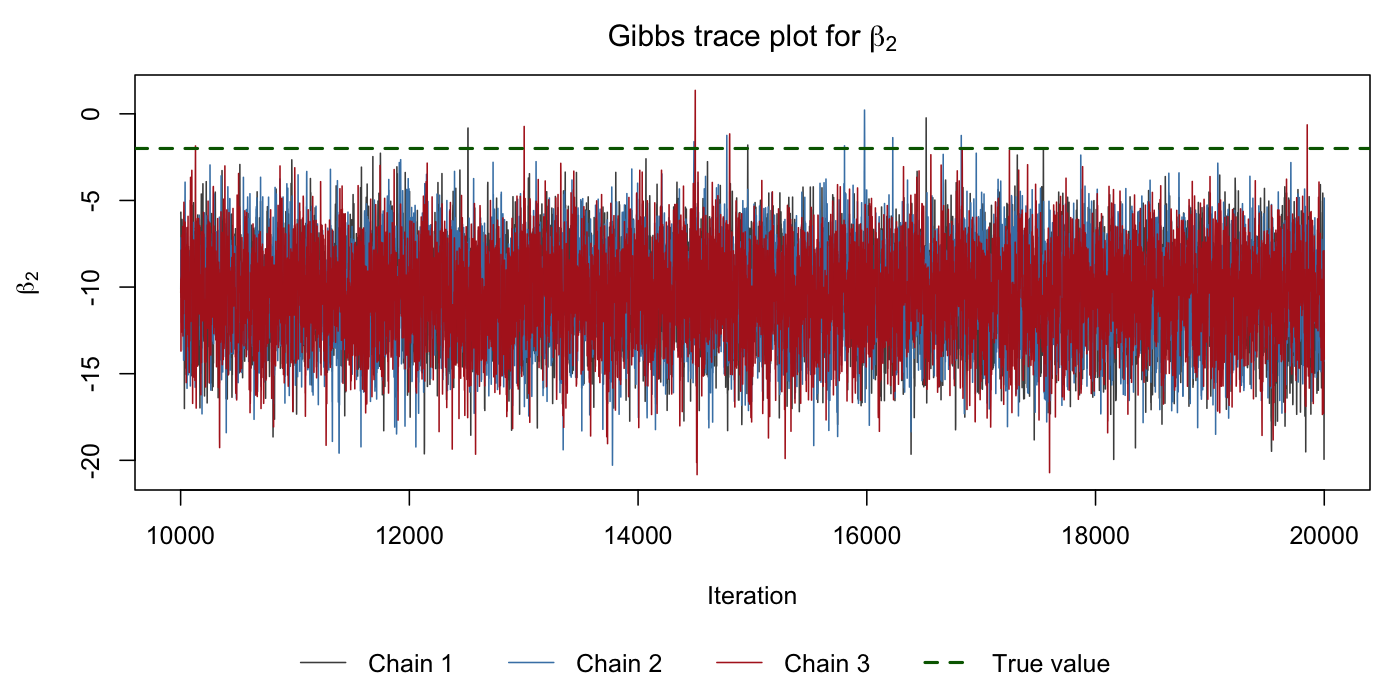}
    \caption{Trace plots for $\beta_2$ after burn-in from three Gibbs chains
    in a representative categorical simulation replicate. Each chain was
    run for 20,000 iterations, with the first 10,000 iterations discarded
    as burn-in. The dashed horizontal line indicates the true value
    $\beta_2=-2$.}
    \label{fig:gibbs-beta2-trace}
\end{figure}

Figure~\ref{fig:gibbs-beta2-trace} shows the trace plots for
$\beta_2$ after burn-in. The three chains show stable behavior and substantial overlap, remaining centered near approximately $-10$ rather than showing systematic drift toward the true value of $-2$. The Gelman--Rubin potential scale reduction factor was $1.00016$ (upper confidence limit $1.00063$), and the effective sample size was approximately $9{,}235$. These diagnostics provide no evidence of substantial lack of convergence for $\beta_2$ and suggest that its poor recovery in the categorical setting is not explained simply by an
insufficient number of Gibbs iterations.

\bibliographystyle{apalike}
\bibliography{references}

\end{document}